\documentclass[twocolumn,fleqn,natbib]{svjour2}
\bibpunct{[}{]}{;}{n}{}{,} 

\smartqed  

\usepackage{graphicx,epsfig,citesort}

\journalname{Granular Matter}

\begin{document}

\title{Influence of particle shape on sheared dense granular media}

\author{A.A. Pe\~na \and R. Garc\'{\i}a-Rojo \and H. J. Herrmann}

\institute{A.A. Pe\~na  \and R. Garc\'{\i}a-Rojo \and H. J. Herrmann
              \at
              Institute for Computer Applications 1,\\
              Pfaffenwaldring 27, \\
              70569 Stuttgart, GERMANY \\
              \email{andres@ica1.uni-stuttgart.de}           
           \and
           H. J. Herrmann \at
              Departamento de Fisica, Universidade Federal do Cear\'a, \\
              Campus do Pici, 60451-970 Fortaleza CE,
              Brazil
}

\date{Received: }

\maketitle

\begin{abstract}
We study by means of  molecular dynamics simulations of periodic
shear cells, the influence of particle shape on the global
mechanical behavior of dense granular media. Results at
macro-mechanical level show that for large shear deformation
samples with elongated particles, independent of their initial
orientation, reach the same stationary value for both shear force
and void ratio. At the micro-mechanical level the stress, the
fabric and the inertia tensors of the particles are used to study
the evolution of the media. In the case of isotropic particles the
direction of the principal axis of the fabric tensor is aligned
with the one of the principal stress, while for elongated
particles the fabric orientation is strongly dependent on the
orientation of the particles. The shear band width is shown to
depend on the particle shape due to the tendency of elongated
particles to preferential orientations and less rotation.

\end{abstract}

\section{Introduction}
\label{introduction}

Granular materials present a complex mechanical response when
subjected to an external load. This global response is strongly
dependent on the discrete character of the medium. The shape,
angularity and size distribution of the grains, the evolution of
the granular skeleton (spatial arrangement of particles, void
ratio, fabric, force chains), and some phenomena occurring at the
grain scale (like rolling or sliding) are determinant factors for
the overall macroscopic response.

The study of anisotropy is of fundamental importance in order to
understand the properties of granular materials. In soil
mechanics, Casagrande and Carrillo \cite{Casag44} distinguished
between inherent and induced anisotropy, the first as a result of
the sedimentation of particles and the second as a product of
inelastic deformation. Oda \cite{Oda72} studied by means of
optical microscopic photographs of thin sections of sand, the
preferred orientations and spatial relation of the constituting
grains, and found that independent of their shape particles tend
to rest in a stable configuration relative to the forces acting on
them. Furthermore, he also found that the initial fabric is not
only determined by the shape of the particles but also by the
deposition or compaction method.

In relation to inherent anisotropy Oda et al. \cite{Oda85} and Oda
and Nakayama \cite{Oda88} have listed three sources in such
materials:

\begin{enumerate}
    \item Anisotropic distribution of contacts (called structural anisotropy).
    \item Shape and preferred orientation of void spaces.
    \item Shape of the particles and preferred orientation of non-spherical ones.
\end{enumerate}

The complete alteration of inherent anisotropy due to types 1 and
2 during early stages of inelastic deformation in biaxial
compression tests, on two-dimensional assemblies of rods, was
observed by Oda et al. \cite{Oda85}. They found, however, that the
one due to type 3 was still present at large deformations. It
would be therefore expected that in the so-called critical state
of soil mechanics, which is associated with large shear
deformation, the persistence of inherent anisotropy is mainly due
to particle orientation \cite{Li02}.

The role of anisotropy in granular materials has been widely
investigated both experimentally and with numerical simulations.
In relation to particle shape, a big effort has been recently done
to properly characterize particle geometry and to study its effect
on the mechanical response of granular media. Concerning particle
shape characterization Bowman et al. \cite{bowman01} proposed a
experimental technique using fourier descriptors and image
analysis to assess particle morphology and texture, and in
discrete element simulations (DEM) Matsushima and Saomoto
\cite{matsuhima02} presented a method to construct
irregularly-shaped grains from shape data of real granular
material. Regarding the influence of particle shape on the
mechanical behavior of the granular media, Bowman and Soga
\cite{bowman05} found that the stress-strain and creep response of
fine silica sand is influenced by particle elongation or aspect
ratio. Using two dimensional DEM simulations Shodja and Nezami
\cite{shodja03} and Nouguier-Lehon and Frossard \cite{nouguier05}
studied the effect of particle elongation on the rolling and
sliding mechanisms in granular media. Nouguier-Lehon et al.
\cite{nouguier03} studied in biaxial test simulations the
influence of particle shape on the so-called critical state, and
Pe\~na et al. \cite{pena05} the effect on the asymptotical states
of granular media. Three dimensional studies using ellipsoidal
convex-shaped particles have also been performed highlighting the
role of particle shape \cite{ng01,ng04}.

The temporal and spatial evolution of anisotropic granular
materials during compaction has also been investigated. Villarruel
et al. \cite{villarruel00} found experimentally clear evidence
that particle anisotropy can drive ordering (configuration of
particles evolves to a nematic state). Lumay and Vandewalle
\cite{lumay04} in experiments and simulations observed large
variations of the asymptotic packing volume fraction as a function
of the aspect ratio of the particles. Ribi\`ere \cite{ribiere05}
using rice of different shapes found that grain anisotropy slows
down the packing fraction evolution, and that the convection in
the granular media and therefore the compaction mechanism depend
also on it.

\begin{figure} [t]
\centering
    \psfig{file=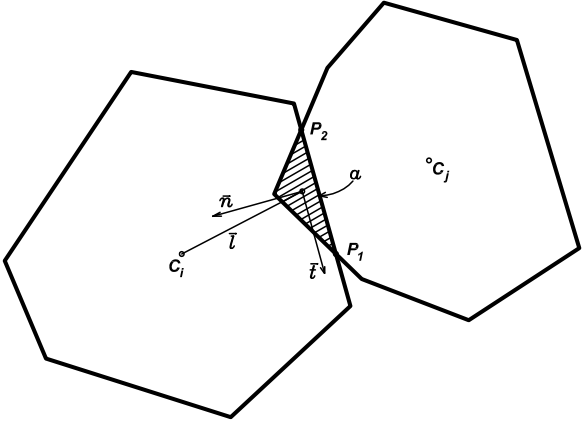,height=6.5cm,angle=0}
    \caption{Schematic representation of a particle contact, the overlapping area $a$ is indicated
    by the shaded zone.}
    \label{Fig:Overlap}
\end{figure}

The anisotropic network of contact forces in a granular packing
subjected to shear conditions has been recently investigated by
Majmudar and Behringer \cite{majmudar05} in a two dimensional
system of photoelastic discs and by L\"atzel et al.
\cite{latzel00} in discrete element simulations.

Up to now, however, there is no clear information at the
micro-mechanical level about the influence of particle shape and
the orientation of non-spherical particles on the evolution of
granular materials and the  corresponding  anisotropic  network of
contacts towards the stationary state reached at the global level
under shear conditions. These subjects are very important in
geotechnical engineering and physics in order to get a better
understanding of the mechanical response of granular materials. In
this work, we use the molecular dynamics technique to simulate the
mechanical behavior of two-dimensional particle assemblies
contained in a periodic shear cell. The grains are represented by
randomly generated convex polygons. We focus on the influence of
particle shape on the overall plastic response. The dependency of
the mechanical behavior on the evolution of inherent anisotropy
(specially contact and non-spherical particle orientations) is
studied. Results are analyzed from the macro and micro-mechanical
point of view.

\section{Model, sample
construction and numerical experiment} \label{Samples}

\subsection{Molecular dynamics simulations}

We study dense polygonal packings, in which particles interact
through visco-elastic contacts. The polygons can neither break nor
deform, but they can overlap when they are pressed against each
other. This overlap represents the local deformation of the
grains. This approach has been frequently used to model many
different processes, such as strain localization and earthquakes
\cite{tillemans95}, fragmentation \cite{kun96b}, and damage
\cite{kun99}.

The repulsive, elastic normal contact force is proportional to the
overlap area $a$ between particles. In Figure $\ref{Fig:Overlap}$
the configuration of a contact between two particles is presented,
$P_1$ and $P_2$ represent the intersection points between the
polygons; the segment that connects those points gives the contact
line $\vec{S} = P_1 P_2$. This vector defines a coordinate system
$(\hat{n},\hat{t})$ at the contact, where $\hat{t}=
\vec{S}/|\vec{S}|$ and $\hat{n}$ normal to it give the direction
of the normal $f_n$ and tangential $f_t$ components of the contact
force. As point of application of the contact forces we choose the
center of mass of the overlapping area \cite{alonso02}, indicated
in the figure with the dot inside the shaded zone.

The elastic force at the contact point is calculated:

\begin{equation}
\vec{f}^c = -k_n \delta \hat{n} - k_t \xi \hat{t}
\label{eq:ContFor}
\end{equation}
where $k_n$ and $k_t$ are the normal and tangential contact
stiffnesses, respectively. The deformation length $\delta$ is calculated in
terms of the overlapping area $a$ and the length of the contact
line $|\vec{S}|$, $\delta = a / |\vec{S}|$. The tangential
friction force is introduced by an elastic spring whose length is
equal to 0 in the case of new contacts, while for old contacts the
tangential spring-length is updated as follows,

\begin{equation}
\xi = \xi' + \vec{v}_t^c \Delta t_{MD}, \label{eq:TangDisp}
\end{equation}
where $\xi'$ is the previous length of the spring, $\Delta t_{MD}$
is the time step of the molecular dynamic simulation, and
$\vec{v}_t^c$ the tangential component of the relative velocity
$\vec{v}^c$ at the contact:

\begin{equation}
\vec{v}^c = \vec{v}_i - \vec{v}_j - \vec{w}_i \times \vec{l}_i +
\vec{w}_j \times \vec{l}_j.
 \label{eq:VeloCt}
\end{equation}
$\vec{v}_i$ is here the linear velocity and $\vec{w}_i$ the
angular velocity of the particles in contact. The tangential
elastic displacement $\xi$ at the contact may increase during the
time that the condition $|f_t^c| < \mu f_n^c$ is satisfied. The
sliding condition is enforced keeping constant the tangential
force when the Coulomb limit condition $|f_t^c| = \mu f_n^c$ is
reached. At this moment the permanent deformation begins.

We also introduce a viscous force at each contact point, which is
necessary to maintain the numerical stability of the method and to
take into account dissipation during particle's contact.

\begin{equation}
\vec{f}_v^c = -m (\nu_n \cdot \vec{v}_n^c \cdot  \hat{n}^c - \nu_t
\cdot \vec{v}_t^c \cdot \hat{t}^c )\label{eq:ViscoF}
\end{equation}
where $m = (1/m_i +1/m_j)^{-1}$ is the effective mass (kg) of the
two particles in contact, and $\nu_n$ and $\nu_t$ are damping
coefficients. Directly related to the restitution coefficients of
the collisions \cite{tillemans95}. A background damping force of
the form $\vec{f}_i^b = -m \nu_b (\vec{v}_i + \vec{w}_i)$ is also
introduced, where $\nu_b$ is the background damping coefficient.
This background damping force is introduced in order to model the friction between the
particles and the bottom (or top) of the shear cell used on the
two-dimensional experiments performed by Veje et. al
\cite{veje99}, and Howell et al. \cite{howell99}.

\subsection{Generation of samples}
\label{Sec.Generation}

\begin{figure} [ht]
\centering
    \psfig{file=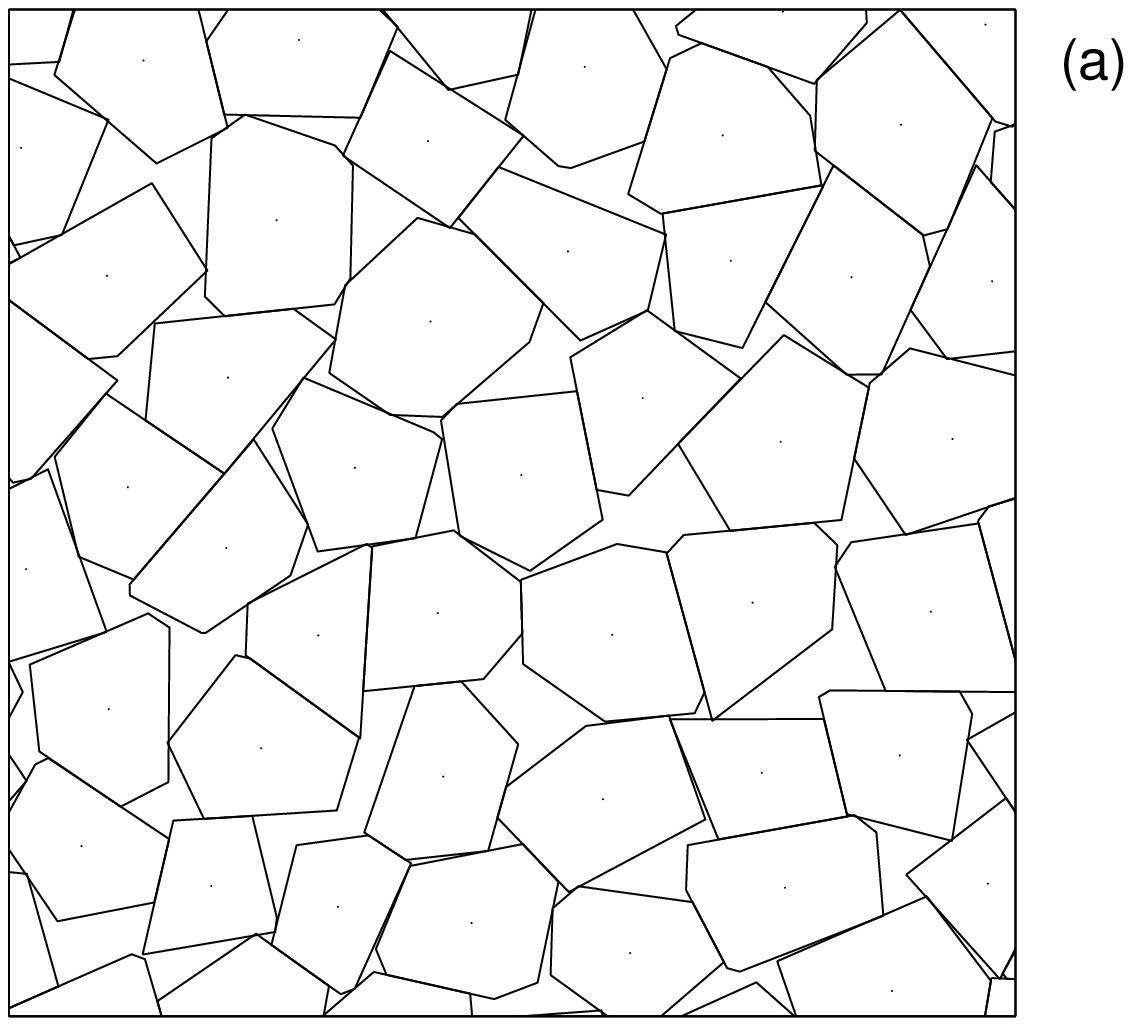,height=4.5cm,angle=0} \\
    \psfig{file=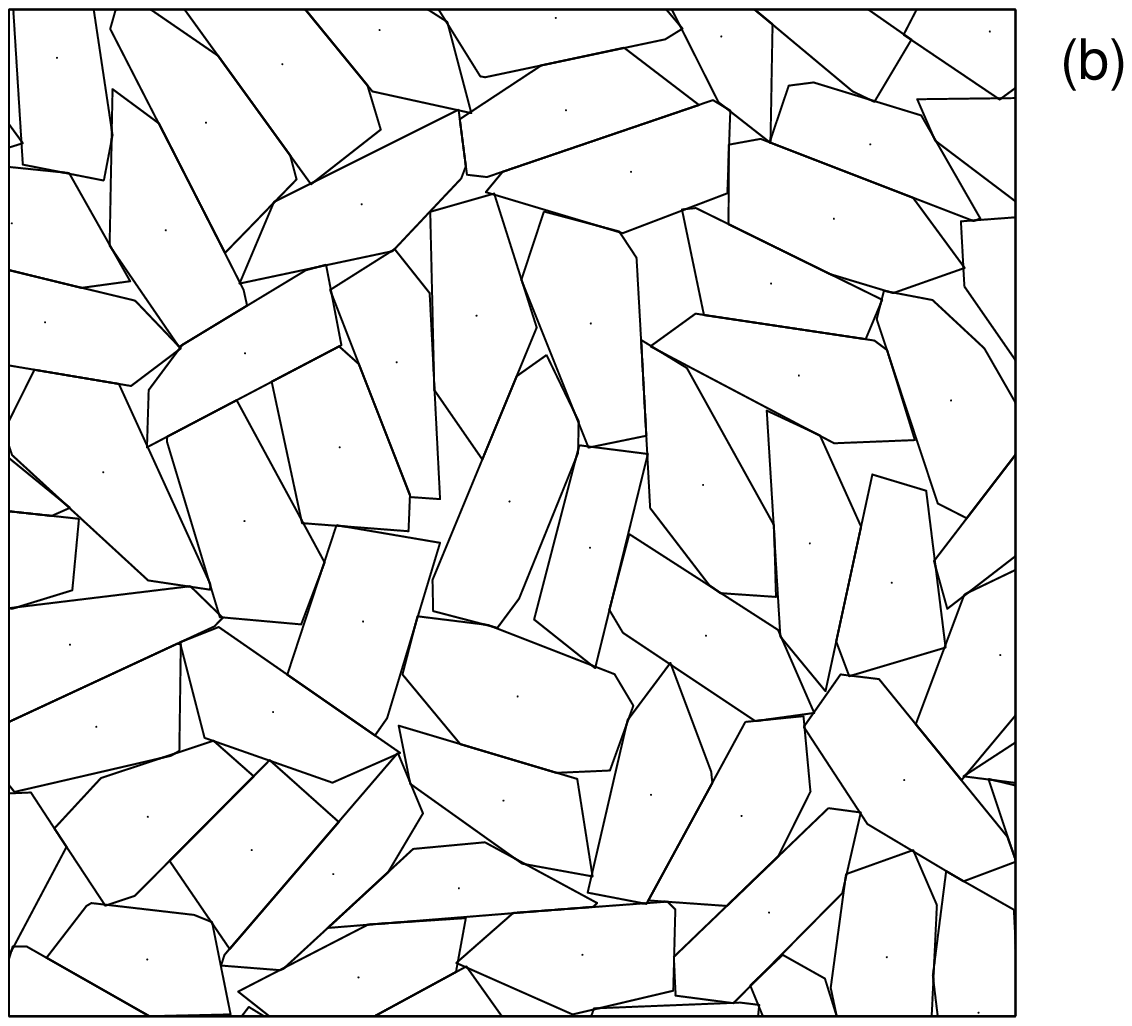,height=4.5cm,angle=0}
    \caption{Two types of particles used in the numerical simulations:
    (a) isotropic $\lambda$ = 1.0 and (b) elongated polygons $\lambda$ = 2.3. Points are the center of mass of the
    polygons.}
    \label{Fig:particles}
\end{figure}

In order to study the influence of particle shape on the
mechanical behavior of granular media, two different types of
convex polygons were used to perform the numerical simulations.
Figure $\ref{Fig:particles}$ shows the visual differences between
the two types of particles. Polygons depicted in Figure
$\ref{Fig:particles}$(a) with an almost isotropic shape, from now
on, will be called 'isotropic' particles, and polygons in
$\ref{Fig:particles}$(b)  will be referred as 'elongated'
particles. The shape of the particles will be characterized by the
aspect ratio $\lambda$, which is obtained from the ratio between
the length of the longest and shortest axis of the particles.

The random generation of the polygons is accomplished by means of
a Voronoi tessellation \cite{moukarzel92}. A regular square
lattice is used to generate the isotropic polygons. This is done
by setting a random point in each cell of the lattice. The
polygons are constructed by assigning to each point the part of
the plane that is nearer to it than to any other point. After
generation, particles are moved apart by multiplying their
coordinates by a constant in order to obtain a very loose state.
Particles are then compressed isotropically by four rigid walls
until the desired confining pressure is reached. The elongated
particles are generated by stretching or contracting, before the
compression, the particles in horizontal or/and vertical direction
by a given factor. The average elongation of the grains is given
by the ratio between these factors $\lambda$ (aspect ratio). In
our numerical simulations we consider two aspect ratios: $\lambda$
= 1.0 for isotropic and $\lambda$ = 2.3 for elongated polygons
(which are clearly anisotropic). Although aspect ratios larger
than 2.3 can be used, we choose this value in order to avoid
particles with very sharp angles that could have an unrealistic
overlap. Nevertheless, this should be verified by means of a
systematic study of different and/or larger aspect ratios of the
particles. Finally, by setting the interparticle friction to zero
during the compression we obtain the dense samples used in our
simulations.

\begin{figure*}[!htb]
\centering
  \psfig{file=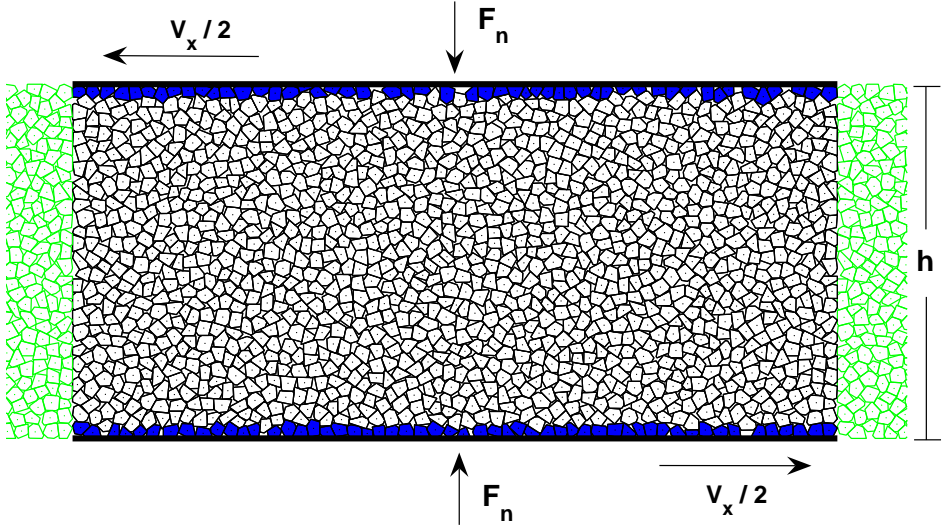,height=7.cm,angle=0}
    \caption{Sketch of the shear cell. A normal force is applied between top and bottom
    wall. A horizontal velocity $v_x$ is fixed shearing the sample. Green (grey) particles correspond
    to the image used to implement the periodic boundary conditions.}
    \label{Fig:SHcell}
\end{figure*}

\subsection{Numerical experiment}

We want to study the behavior of the granular media in a periodic
shear cell. The cell contains 1500 particles, being 50 particle
diameters wide and 30 diameters high. Periodic boundary conditions
are imposed in horizontal direction. The top and bottom have fixed
boundary conditions. A constant confining stress (or normal force)
is imposed between the bottom and the top horizontal walls. The top
and bottom layers of particles are sheared in opposite direction
with a fixed horizontal velocity $v_x$ relative to each other. The
particles in one layer are not allowed to rotate or move against
each other. The top boundary is free to move in vertical direction
in order to permit a volumetric change of the sample, while the
bottom is kept fixed. A configuration of the shear cell used in
our simulations is depicted in Figure $\ref{Fig:SHcell}$.

In all simulations the mechanical parameters of the particles are:
interparticle friction coefficient $\mu=0.5$, normal stiffness at
the contact $k_n$ = $1.6 \cdot 10^8$ N/m, and normal damping
coefficient $\nu_n$ = 4000  s$^{-1}$. The stiffness ratio
$\zeta=k_t/k_n$ as well as the viscosity ratio $\nu_t/\nu_n$ were
taken to be 1/3. A background damping coefficient $\nu_b$ = 12
s$^{-1}$ was used. Further simulations, results not presented
here, with no and different background damping coefficients $\nu_b$ in order to
evaluate its influence on the mechanical behavior of the medium
were performed. We found that the damping $\nu_b$ in the range
used here has no effect on the evolution of the internal variables
of the medium, and additionally that the $\nu_b$ value used in
this work has no influence on the global mechanical response
either.

The horizontal and vertical directions are indicated as $x$ and
$y$, respectively. In order to study the evolution of the packing
we use the strain variable $\gamma$, which is defined as follows:

\begin{equation}
\gamma =  D_x / h_o,
\end{equation}
where $D_x$ is the horizontal displacement of the boundary
particles and $h_o$ is the initial height of the sample. The void
ratio $e$ of the sample is related to the volumetric deformation,
and is calculated:

\begin{equation}
\ e =  V_T  /  V_S - 1,
\end{equation}
where $V_T$ is the total volume of the sample and $V_S$ the volume
occupied by all the particles.

\section{Shear test simulations}
\label{Experiment}

\subsection{Global mechanical behavior (effect of initial configuration).}

\subsubsection{Statistically different samples}

\begin{figure} [t]
\centering
    \psfig{file=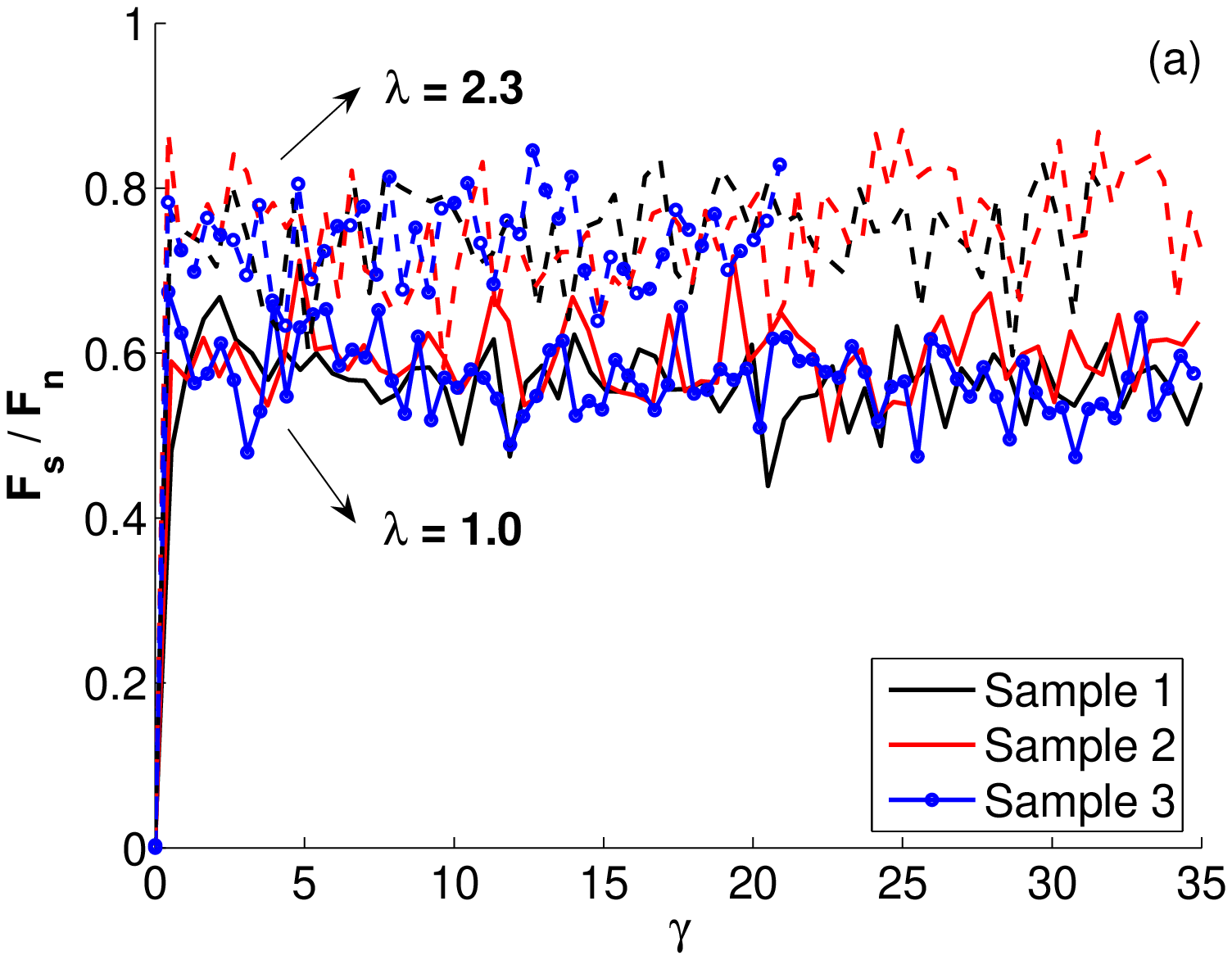,height=5.5cm,angle=0} \\
    \psfig{file=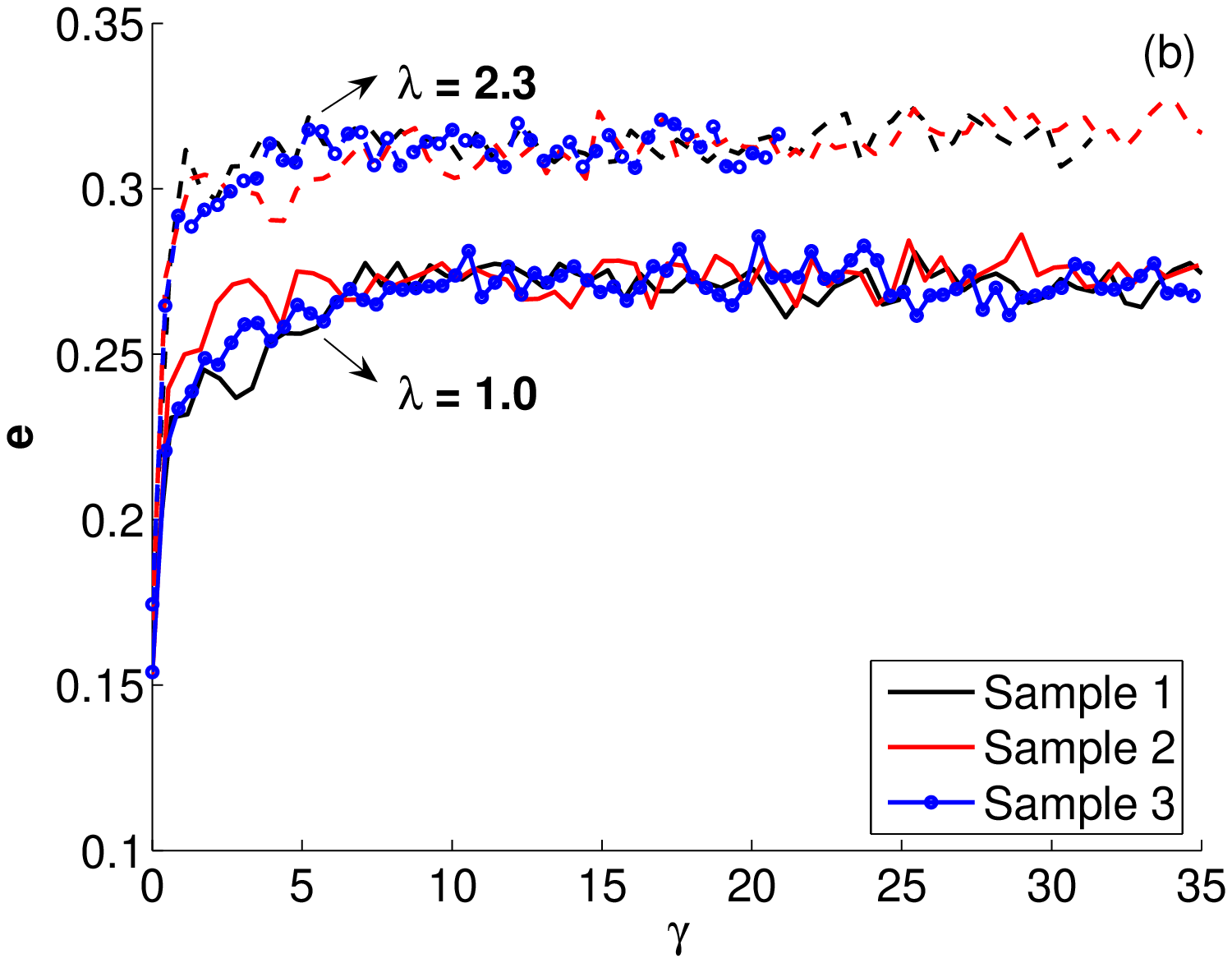,height=5.5cm,angle=0}
    \caption{Evolution of (a) shear force and (b) void ratio for
    different samples with the same mechanical parameters.
    Isotropic ($\lambda$ = 1.0) and elongated particles ($\lambda$ = 2.3) are represented
    by solid and dashed lines, respectively.} \label{Fig:Seed}
\end{figure}

Samples corresponding to different seeds for the random number
generation of the Voronoi tesellation are used to evaluate the
global mechanical response of the granular packing. This is done
in order to assess whether different initial configurations of
particles reach the same steady state. Our results correspond to a
shear velocity of 10 cm/s, and to elongated polygons initially
oriented perpendicular to shear direction. In Figure
$\ref{Fig:Seed}$ the evolution of the resultant shear force and
the void ratio is presented for the different configurations. In
Figure $\ref{Fig:Seed}$(a), the shear force $F_s$ is normalized by
the normal force $F_n$ applied to the system. Initially, the ratio
$F_s/F_n$ has a strong increment related to the breaking of the
interlocking of the particles. After this stage, a saturation
towards a nearly constant value of the $F_s/F_n$ ratio necessary
to shear the granular media is observed, this behavior is
identical for all the samples. For small values of strain, the
evolution of the void ratio (Figure $\ref{Fig:Seed}$b) also
presents a high initial increase saturating later at a constant
value. This saturation occurs slower than for $F_s/F_n$. Samples
with elongated particles saturate at a higher value of $F_s/F_n$
and also higher void ratio, as a consequence of the stronger
interlocking due to the particle shape. One can therefore conclude
that elongated grains are more sensitive to volumetric changes and
develop a higher shear strength. This result had been in fact
previously observed \cite{nouguier03,pena05}. We consider this
saturation of the $F_s/F_n$ value and void ratio $e$ as the steady
state of the sheared material.

\begin{figure} [!h]
\centering
    \psfig{file=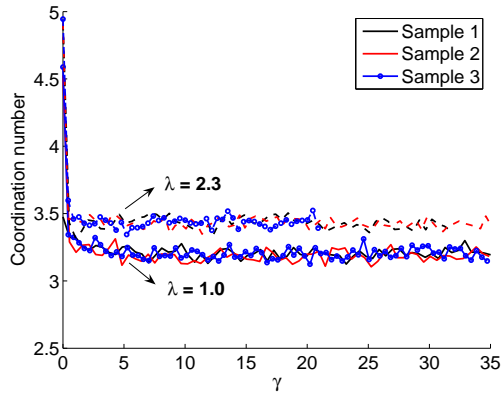,height=5.5cm,angle=0}
    \caption{Evolution of the coordination number for different samples with the same
    mechanical parameters. Isotropic particles $\lambda$ = 1.0 (solid lines) and elongated ones
    $\lambda$ = 2.3 (dashed lines).}
    \label{Fig:CooNumber}
\end{figure}

\begin{figure} [ht]
\centering
    \psfig{file=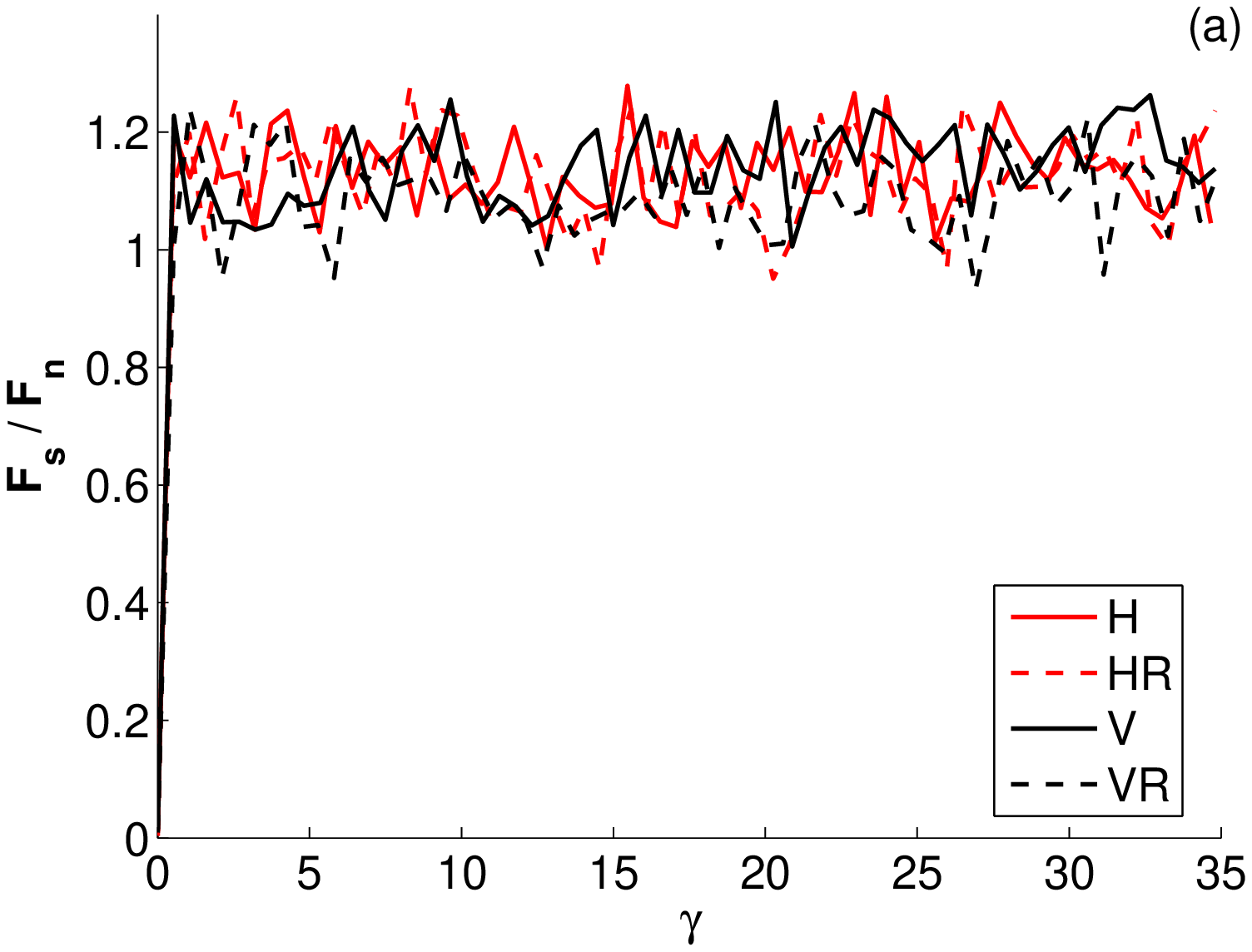,height=5.5cm,angle=0}
    \psfig{file=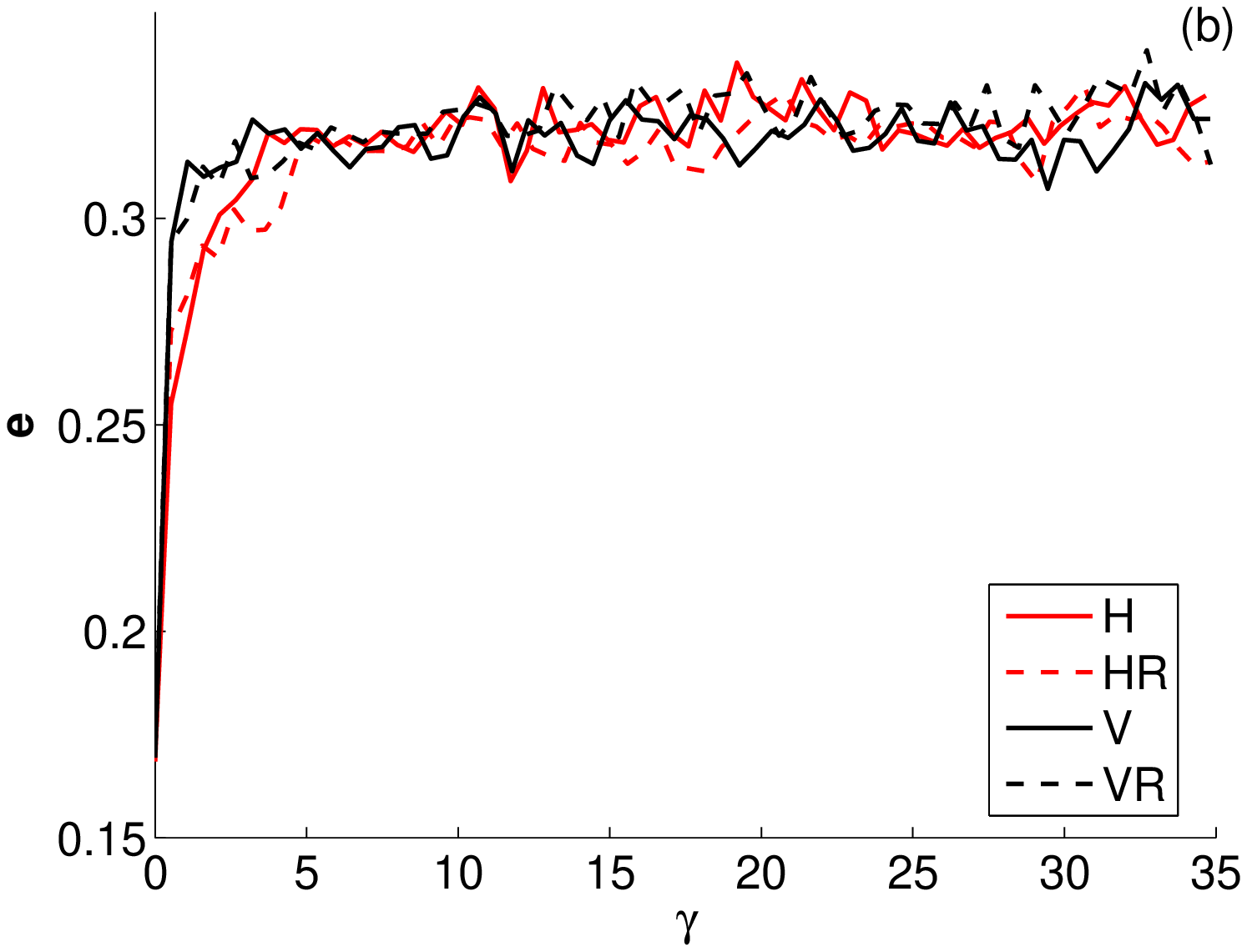,height=5.5cm,angle=0}
    \caption{Evolution of (a) shear force and (b) void ratio for
    samples with different initial particle orientation, $\lambda$ = 2.3. Samples
    labelled H and V have particles oriented in horizontal and vertical
    direction, respectively. R corresponds to an initial random rotation
    of the particles.}
\label{Fig:MBIorient}
\end{figure}

The evolution of the coordination number for isotropic and elongated
particles is depicted in Figure $\ref{Fig:CooNumber}$.
Note that, despite of reaching a higher void ratio, samples with
elongated particles saturate at larger value of coordination
number compared to isotropic particles. This can be understood in
terms of a geometrical effect, consequence of the flat shape
and/or larger relative plane surface in the $\lambda$ = 2.3 case,
which allows for a higher number of contacts per particle.

\subsubsection{Different initial particle orientations}
\label{InitialCon}

We study in this Section the influence of the anisotropy on the
macroscopic behavior of granular media due to the initial
orientation of elongated particles. Three different initial
configurations are obtained for the samples used in this analysis:

\begin{enumerate}
    \item On average the grains are oriented parallel to the shear
    direction (this will be called "horizontal" sample - H).
    \item On average the grains are oriented perpendicular to the shear
    direction (we will call this the "vertical" sample - V).
    \item Grains (H or V) are randomly rotated before isotropic
    compression (which we call the "random" samples - HR or VR).
\end{enumerate}

Configurations number 1 and 2 correspond to samples with different
initial orientation of the particles. Configuration number 3 is
equal to configurations 1 and 2, but with an additional induced
random rotation to the particles before the compression (between 0
and $2\pi\;rad$). In all three cases the samples wind up having a
slight deviation from the originally induced anisotropy due to
particle interactions during the compression.

In Figure $\ref{Fig:MBIorient}$, the evolution of $F_s/F_n$ and
the void ratio $e$ for samples with particles initially oriented
in horizontal and vertical direction, and an additional random
rotation is presented. Results correspond to a shear velocity of
40 cm/s. We notice that $F_s/F_n$ and the void ratio $e$ evolve
toward the same saturation value when they reach the steady state
independently of the initial anisotropy due to contact and
particle orientations. This independence of the initial anisotropy
will be explained by studying the evolution of the internal
variables in Section $\ref{Sec.Int.varib}$.

\subsection{Evolution of internal variables}
\label{Sec.Int.varib}

In order to get a better understanding of what is occurring at the
micro-mechanical level and to find some explanations for the
macro-mechanical behavior observed in the previous section, we
study the evolution of the local stress, the fabric and the
inertia tensors of the samples. Three samples are used: one with
isotropic polygons, and two with elongated polygons (corresponding
to an initial horizontal and vertical orientation of the
particles, as explained in the previous section). Before showing
these results, some key concepts that will be required for the
description of the media are introduced.

\subsubsection{Definitions}

The anisotropy of the contact network within the granular sample
can be characterized by the fabric tensor of second order
$\mathbf{F}$. The fabric tensor takes into account the
distribution of the orientations of the contacts between
particles, that is to say the geometrical structure of the medium
\cite{Oda85}. For a single particle $p$ its components $F_{ij}$
are obtained from

\begin{equation}
F_{ij}^p = \sum_{c=1}^{C^p} l_i^c  l_j^c
\label{eq:fabricTp}
\end{equation}
where the dyadic product of the so-called branch vector $l^c$,
connecting the center of mass of the particle to the contact point
$c$, is summed over all the contacts $C_p$ of particle $p$. Note
that, so defined, the trace of the fabric tensor $F_{ii}^p$ is
then the number of contacts $C_p$ of the particle $p$. It is also
possible to define a normalized fabric tensor $F_{ij}^p$/$C_p$,
whose trace is unity. Finally, the mean fabric tensor for an
assembly can then be defined as follows:

\begin{equation}
F_{ij} = \frac{1}{N_p} \sum_{p=1}^{N_p} F_{ij}^p
\label{eq:fabricT}
\end{equation}
where the particle fabric tensor $F_{ij}^p$ is summed over the
total number of particles $N_p$ within a representative volume
element (RVE). The trace of this tensor is the local mean
coordination number $C_m$, and therefore the normalized mean
fabric tensor can also be defined as $F_{ij}$/$C_m$.

The inertia tensor can be calculated for each particle as follows:

\begin{equation}
i_{ij}^p = \int \rho (\delta_{ij} \sum_{k} x_k^2 - x_i x_j) dA
\label{eq:inertiaT}
\end{equation}
where $\rho$ is the density of the particles (kg/m$^2$),
$\delta_{ij}$ is the Kronecker delta symbol, $k$ runs in our
two-dimensional case from 1 to 2, $dA$ is the differential area
element, and $x$ is the shortest distance from the rotation axis
to $dA$.

The mean inertia tensor is calculated in terms of
($\ref{eq:inertiaT}$):

\begin{equation}
I_{ij} = \frac{1}{N_p} \sum_{p=1}^{N_p} i_{ij}^p \label{eq:Tinert}
\end{equation}
where the particle inertia tensor $i_{ij}^p$ is summed over the
total number of particles $N_p$ in the RVE.

We also calculate the stress tensor, which is defined in terms of
the contact force $f_j^c$ between the grains (acting at the
contact point $c$), and the branch vector $l_i^c$ belonging to the
contact point. The stress tensor is defined as follows
\cite{rothenburg81}:

\begin{equation}
\sigma_{ij} = \frac{1}{V} \sum_{c=1}^{N_c} l_i^c f_j^c
\label{eq:stressT}
\end{equation}
where $V$ is the volume of the RVE, and the summatory extends over
all the contacts $N_c$ in the RVE.

Finally, we will compute the principal directions (major $M$ and
minor $m$) of the mean fabric $\mathbf{F}$, inertia $\mathbf{I}$
and stress $\mathbf{\sigma}$ tensors. These principal
directions are measured with the horizontal axis $x$. We denote
$\theta_F$ the major principal direction of the fabric tensor,
$\theta_I$ the major principal direction of the inertia tensor and
$\theta_{\sigma}$ the one of the stress tensor. We are also
interested in the individual orientation of non-isotropic
particles. We designate $\theta^p$ the angle made by the major
principal direction of the inertia tensor $i_M^p$ of the particle
$p$ with the horizontal axis $x$. Therefore, $\theta^p$ gives the
preferred orientation for each particle of the assembly

\subsubsection{Evolution of the internal variables}

\begin{figure*} [!tbh]
\centering
    \psfig{file=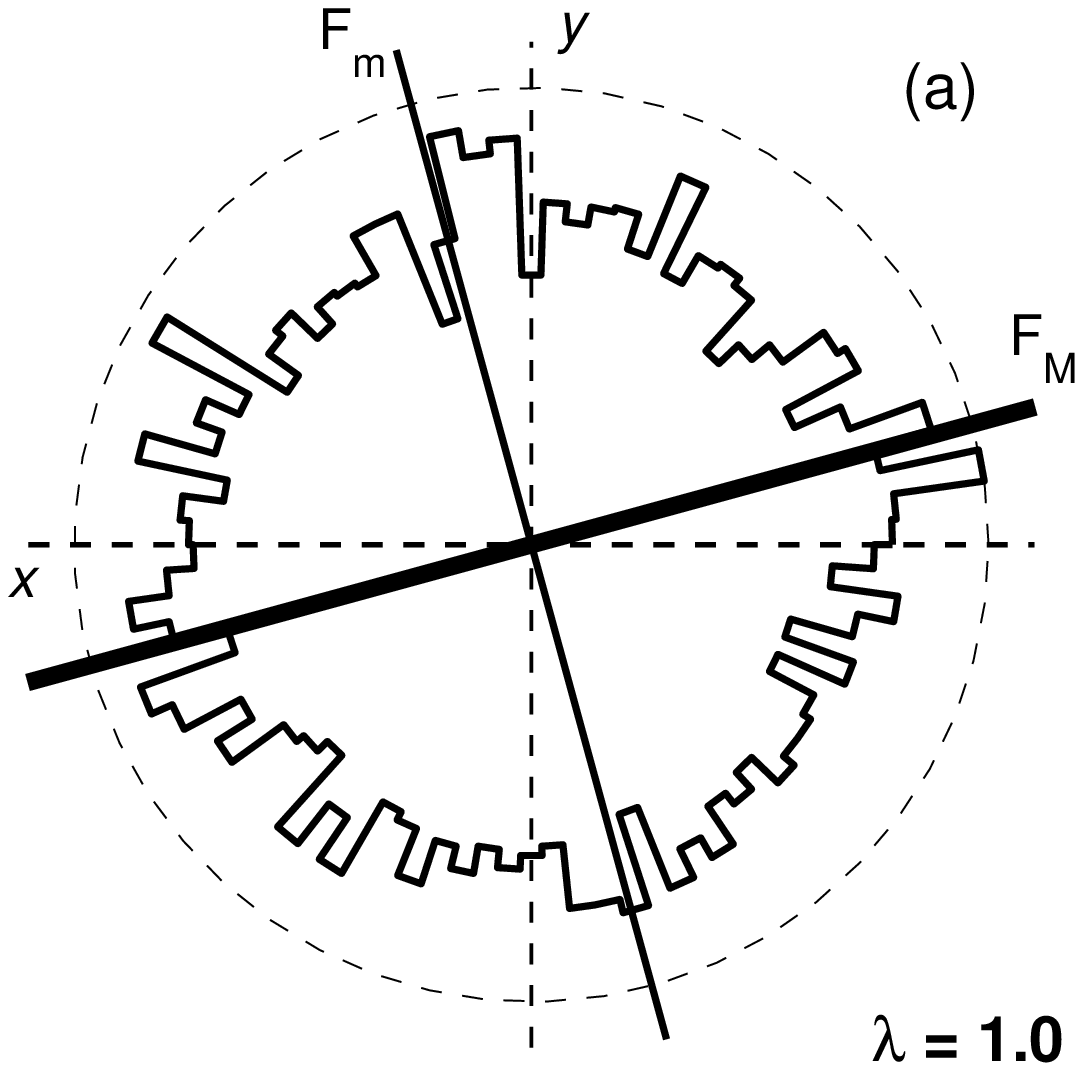,height=4.6cm,angle=0}
    \hspace{1.0cm}
    \psfig{file=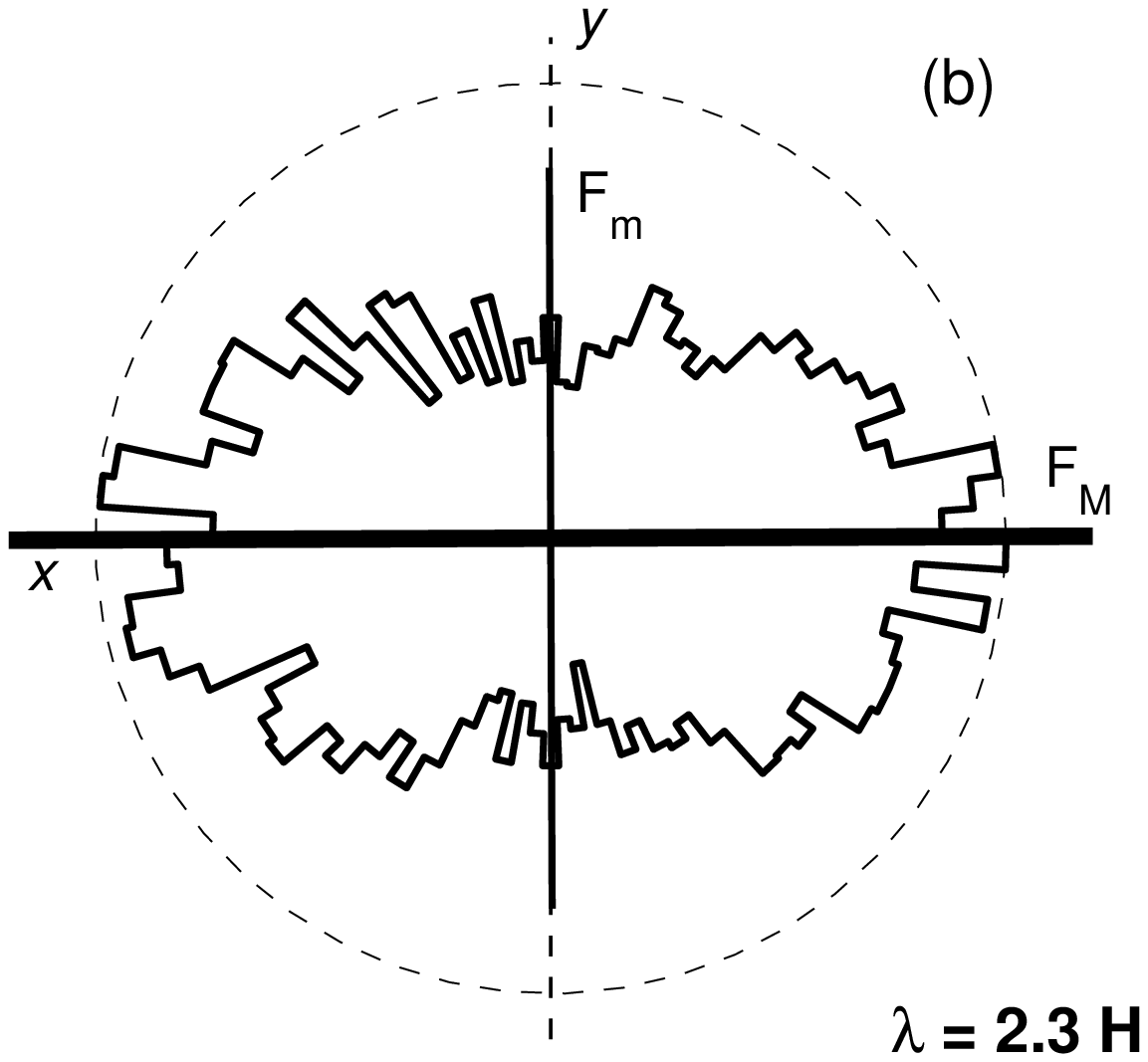,height=4.6cm,angle=0}
    \hspace{1.0cm}
    \psfig{file=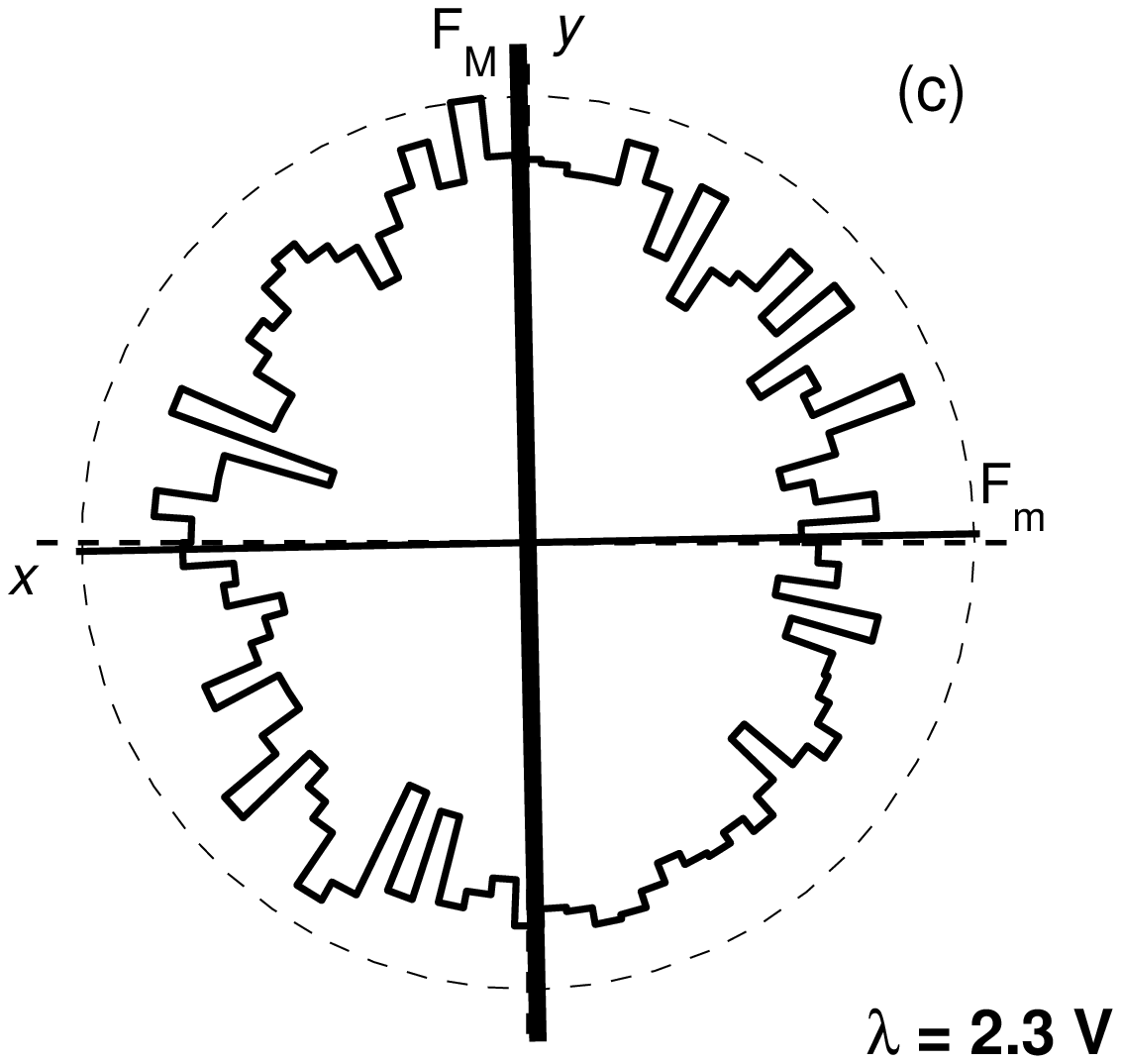,height=4.6cm,angle=0} \\
    \psfig{file=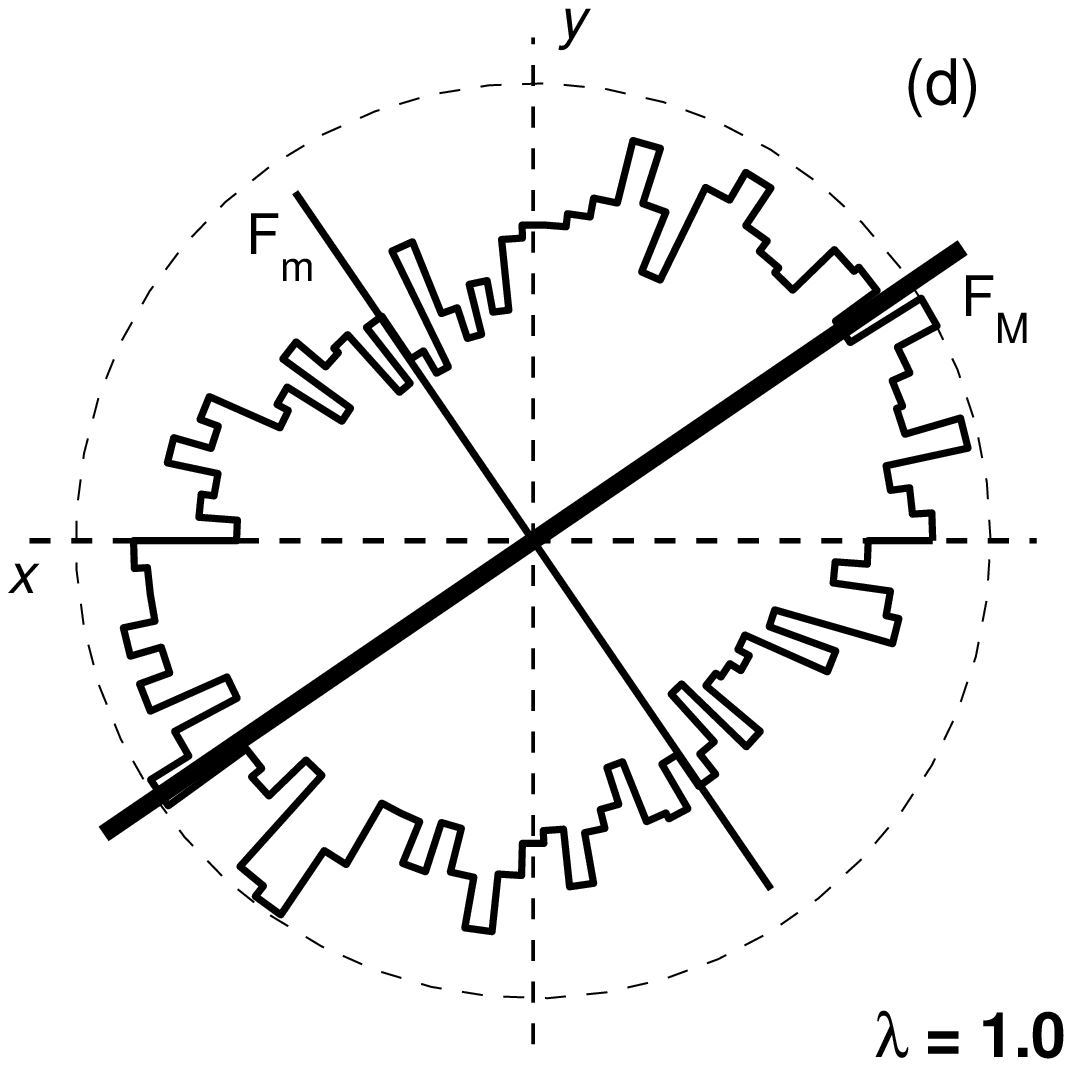,height=4.6cm,angle=0}
    \hspace{1.0cm}
    \psfig{file=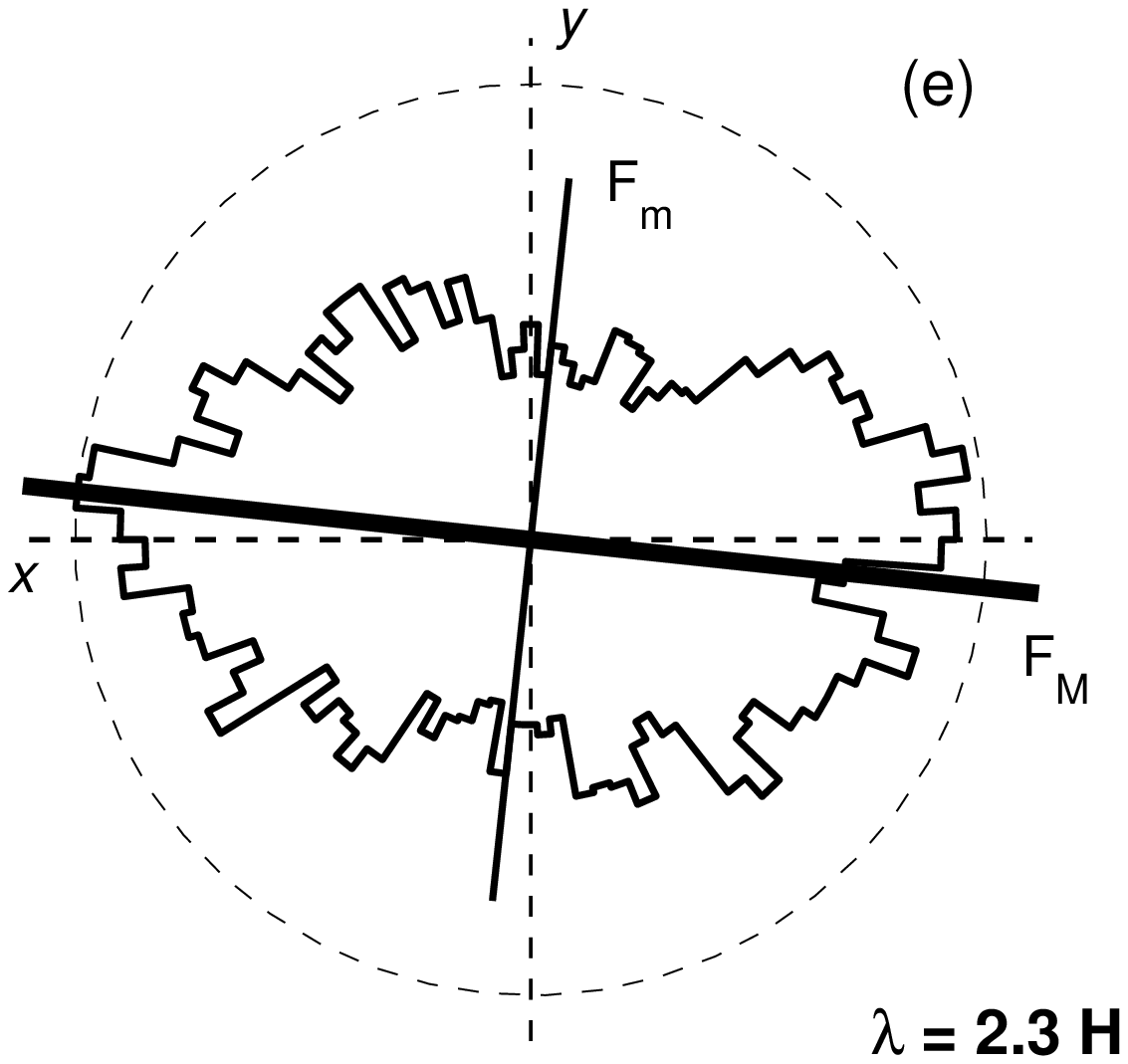,height=4.6cm,angle=0}
    \hspace{1.0cm}
    \psfig{file=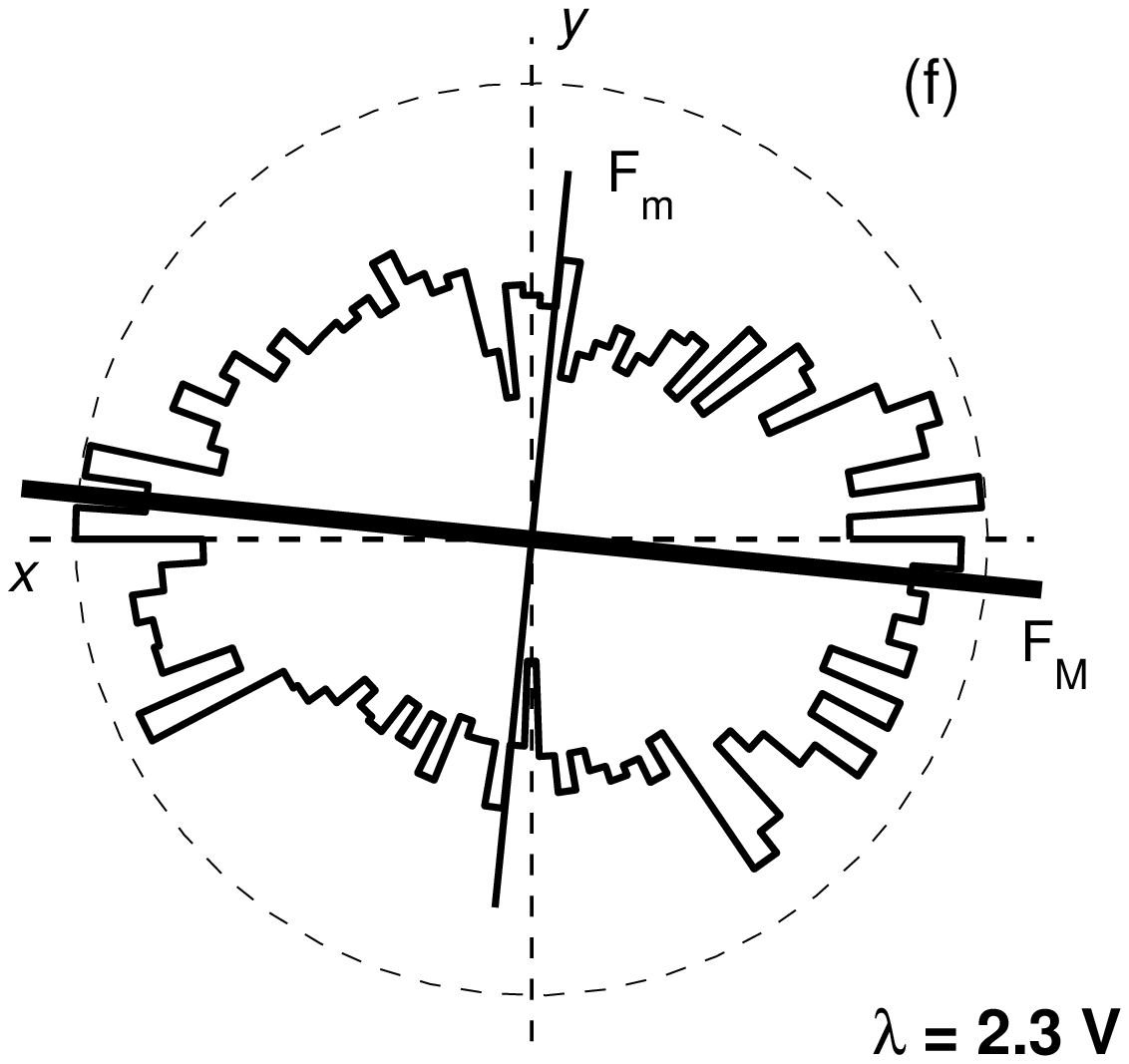,height=4.6cm,angle=0}
    \caption{Polar distribution of branch vectors in the initial
    configuration (a,b,c) and the steady state (d,e,f),
    for isotropic particles  $\lambda$ = 1.0 (a,d), and elongated particles ($\lambda$ = 2.3) initially oriented in horizontal
    direction (b,e) and in vertical direction (c,f). The principal directions of the mean fabric
    tensor ($F_M$  and  $F_m$), and the reference axes $x$ and $y$ are plotted with solid and dashed
    lines, respectively. The radius of the dashed circle corresponds to the maximum value of the distribution. }
\label{Fig:Init-Steady-fabric}
\end{figure*}

In Figure $\ref{Fig:Init-Steady-fabric}$(a-c), we show the orientational
distribution of the branch vectors and the principal directions of
the mean fabric tensor ($F_M$  and  $F_m$) for the initial
configuration of the samples. Observe that, in the case of isotropic polygons
(Fig. $\ref{Fig:Init-Steady-fabric}$(a)), the distribution presents no
preferred direction within the statistical fluctuations. For elongated polygons
(Fig. $\ref{Fig:Init-Steady-fabric}$(b-c)), however, one can
observe that the major principal component of the fabric tensor
$\mathbf{F}$ is oriented towards the direction in which the
polygons were initially stretched.

\begin{figure*} [!hbt]
\centering
    \psfig{file=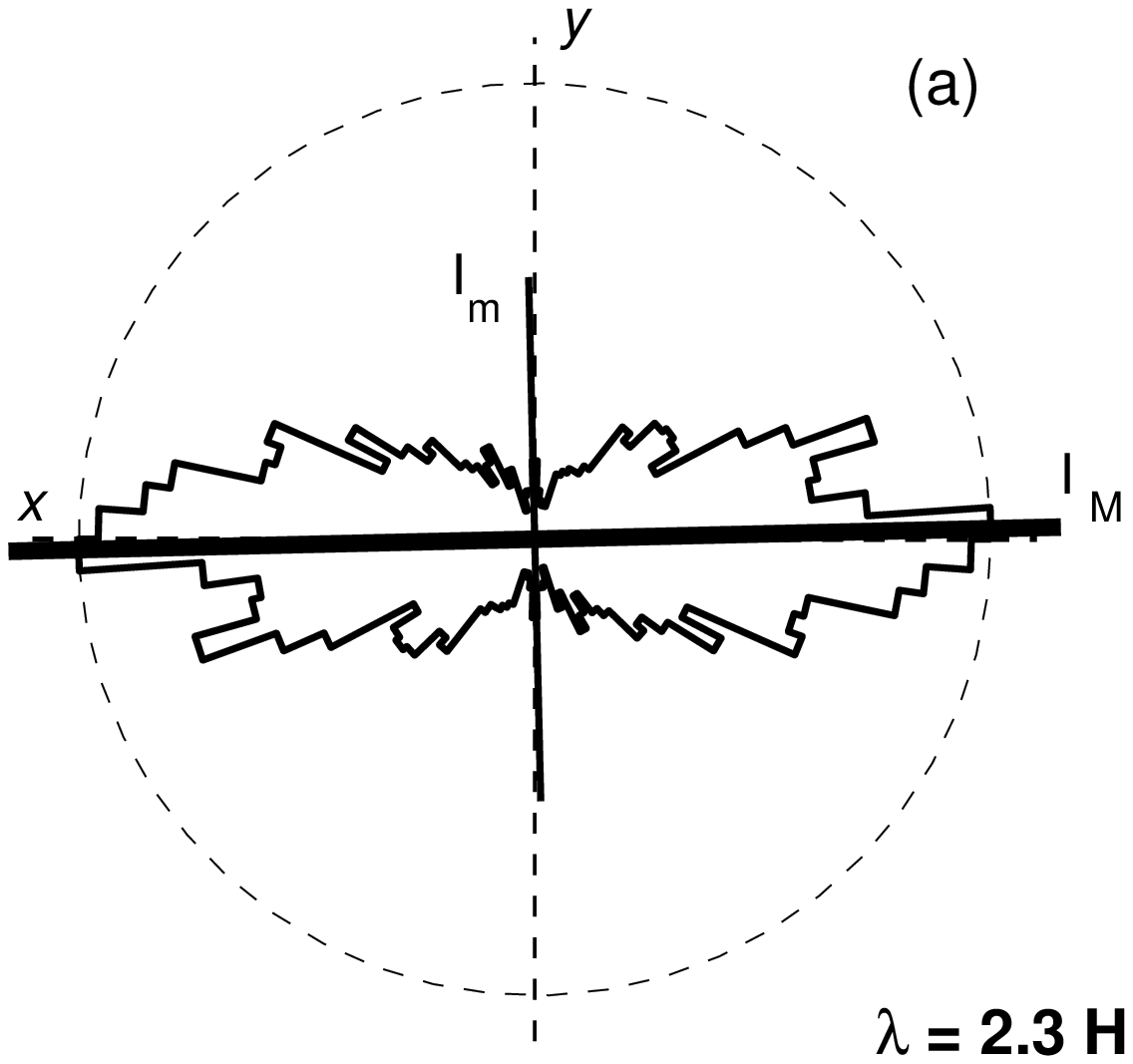,height=4.6cm,angle=0}
    \hspace{2.0cm}
    \psfig{file=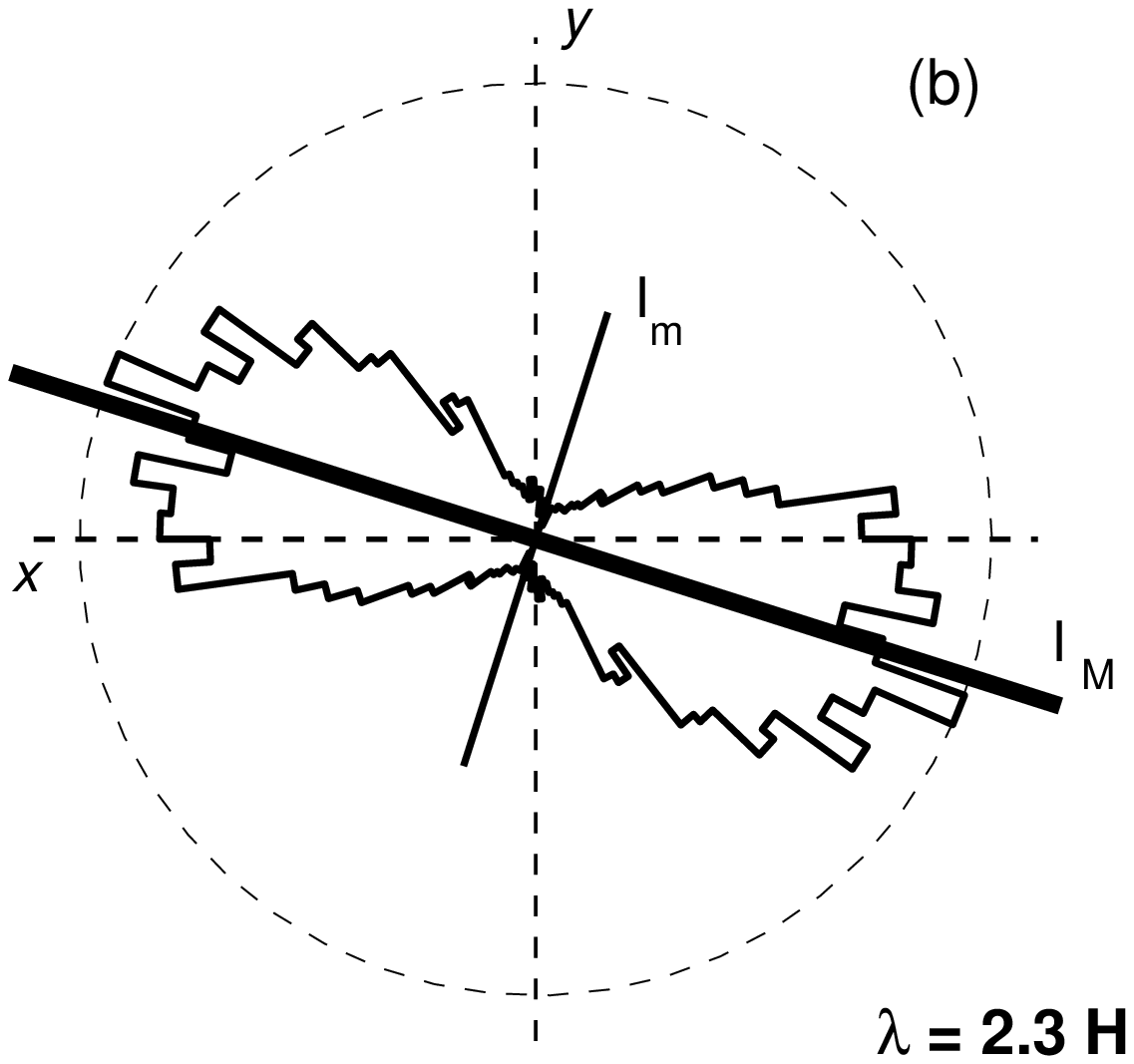,height=4.6cm,angle=0}  \\
    \psfig{file=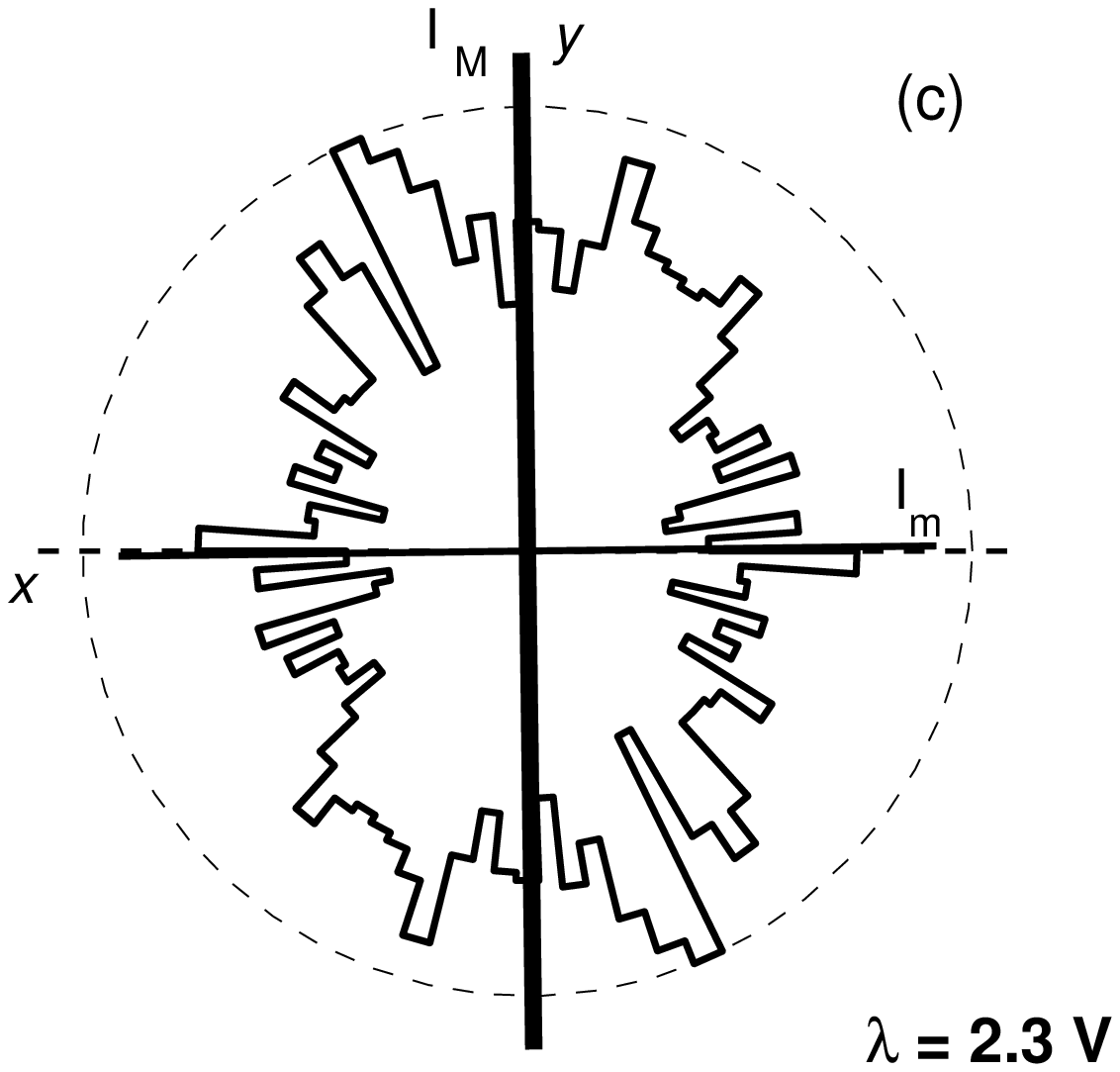,height=4.6cm,angle=0}
    \hspace{2.0cm}
    \psfig{file=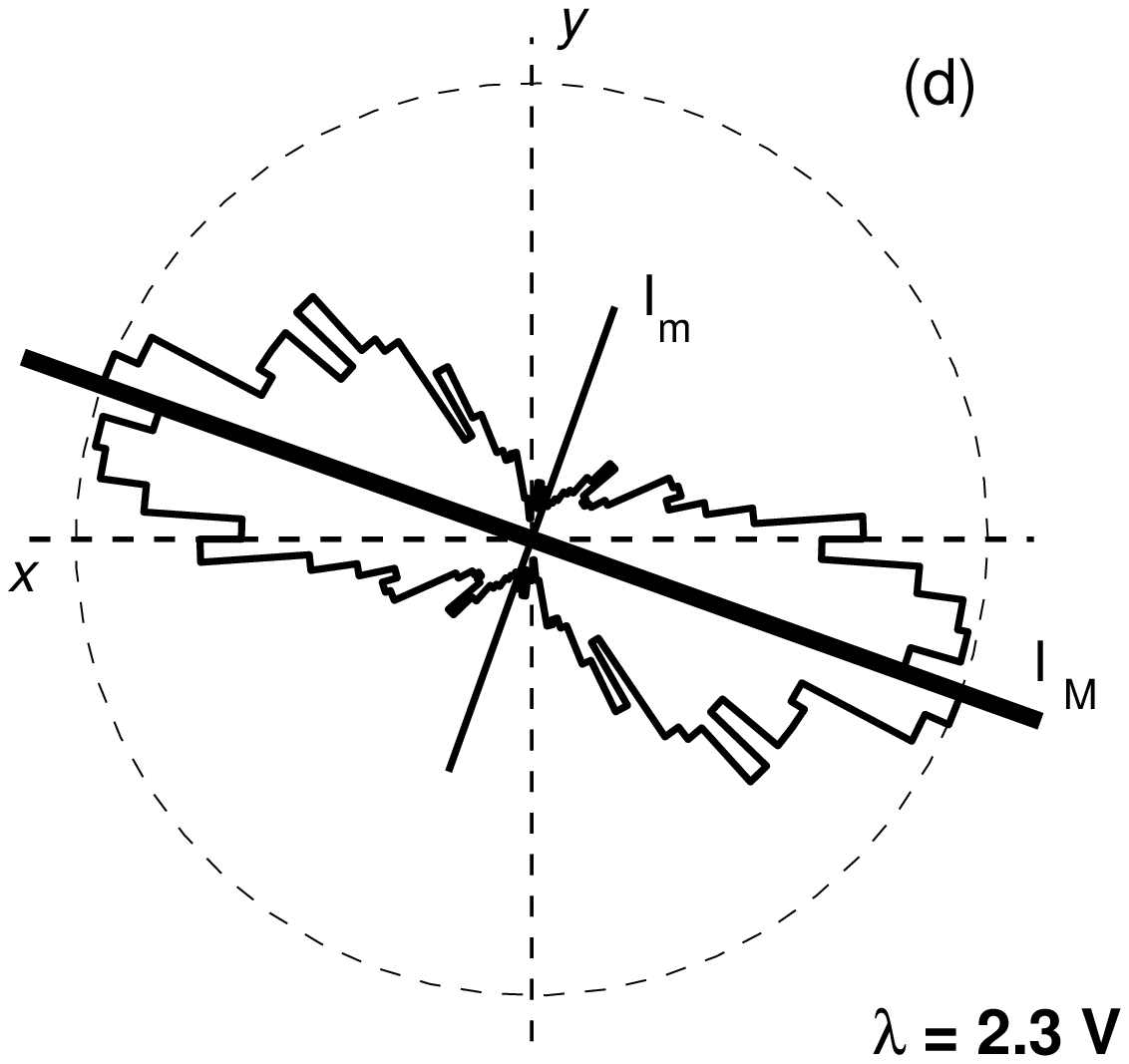,height=4.6cm,angle=0}
    \caption{Polar distribution of particle orientations $\theta^p$, initial configuration (a,c) and
    in the steady state (b,d) for elongated particles ($\lambda$ = 2.3) initially oriented in horizontal
    direction (a,b) and in vertical direction (c,d). The principal directions of the
    global inertia tensor ($I_M$  and  $I_m$), and the reference axes $x$ and $y$ are plotted with solid and dashed
    lines, respectively. The radius of the dashed circle corresponds to the maximum value of the distribution}
\label{Fig:ParticleOri}
\end{figure*}

\begin{figure}
\centering
    \psfig{file=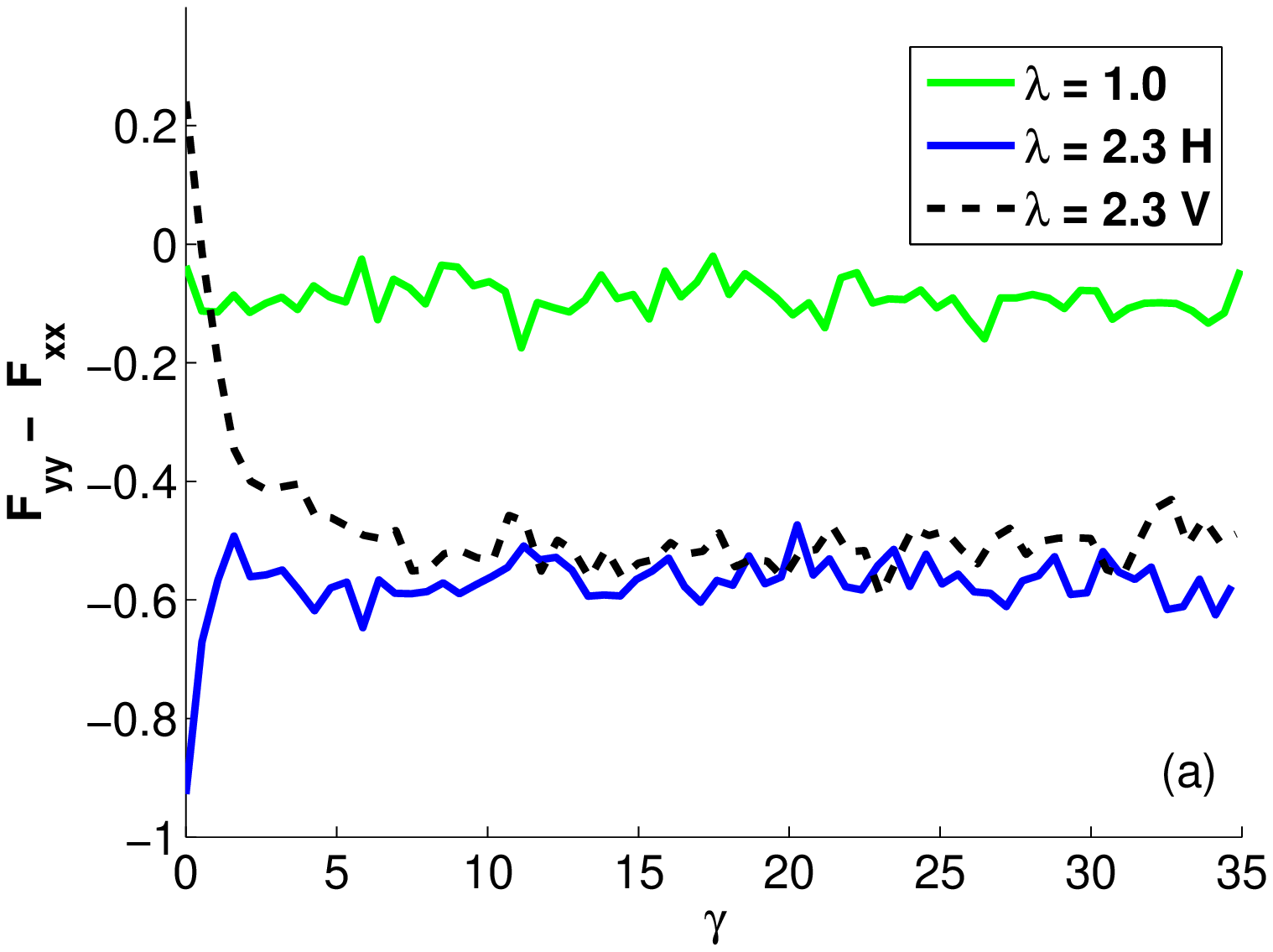,height=5.3cm,angle=0} \\
    \psfig{file=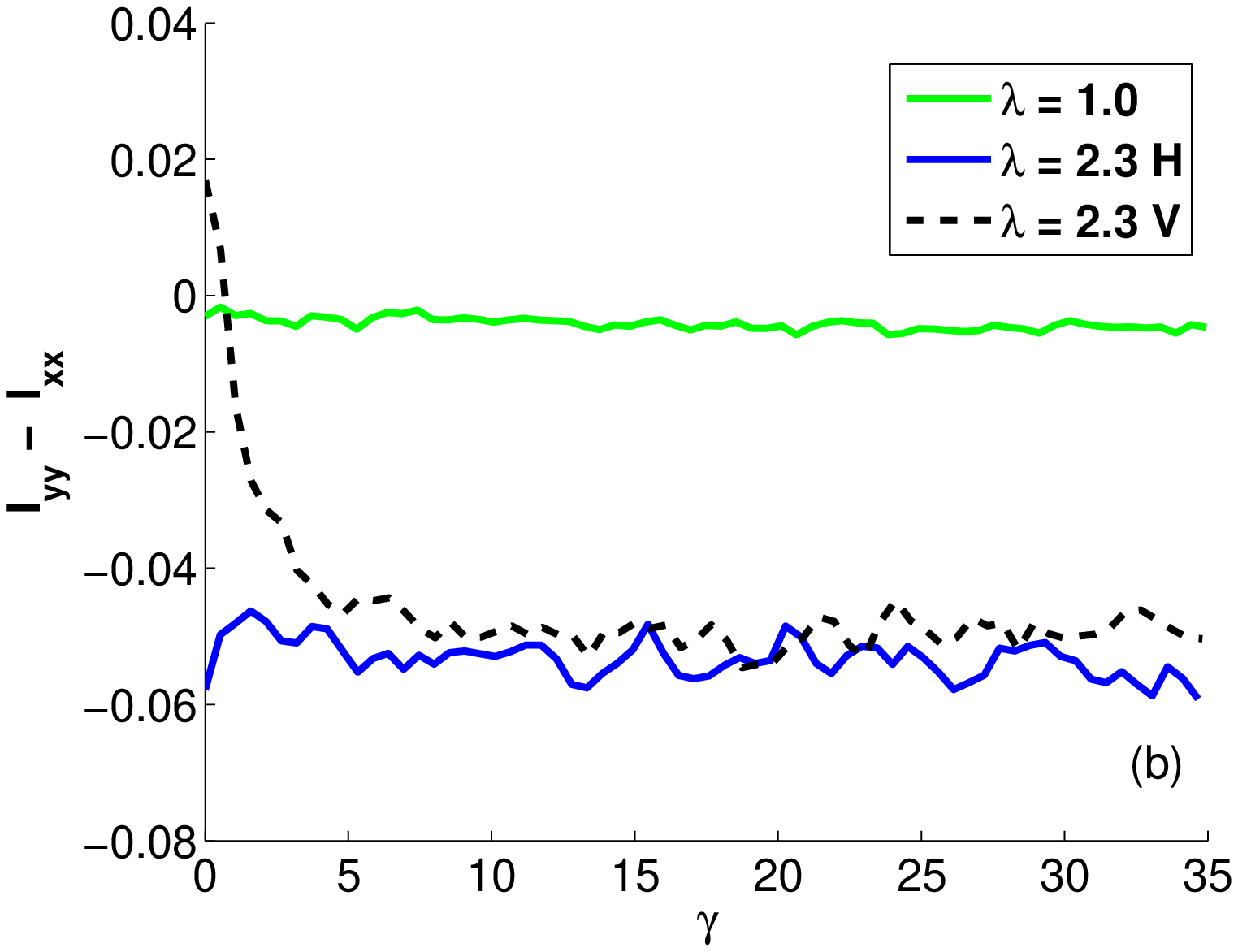,height=5.3cm,angle=0}
    \caption{Evolution of the deviatoric component of the fabric (a) and the
    inertia tensor (b)  for isotropic particles  $\lambda$ = 1.0, and elongated particles ($\lambda$ =
    2.3)
    initially oriented in horizontal direction (H) and in vertical direction (V). } \label{Fig:DevCompon}
\end{figure}

\begin{figure}
\centering
    \psfig{file=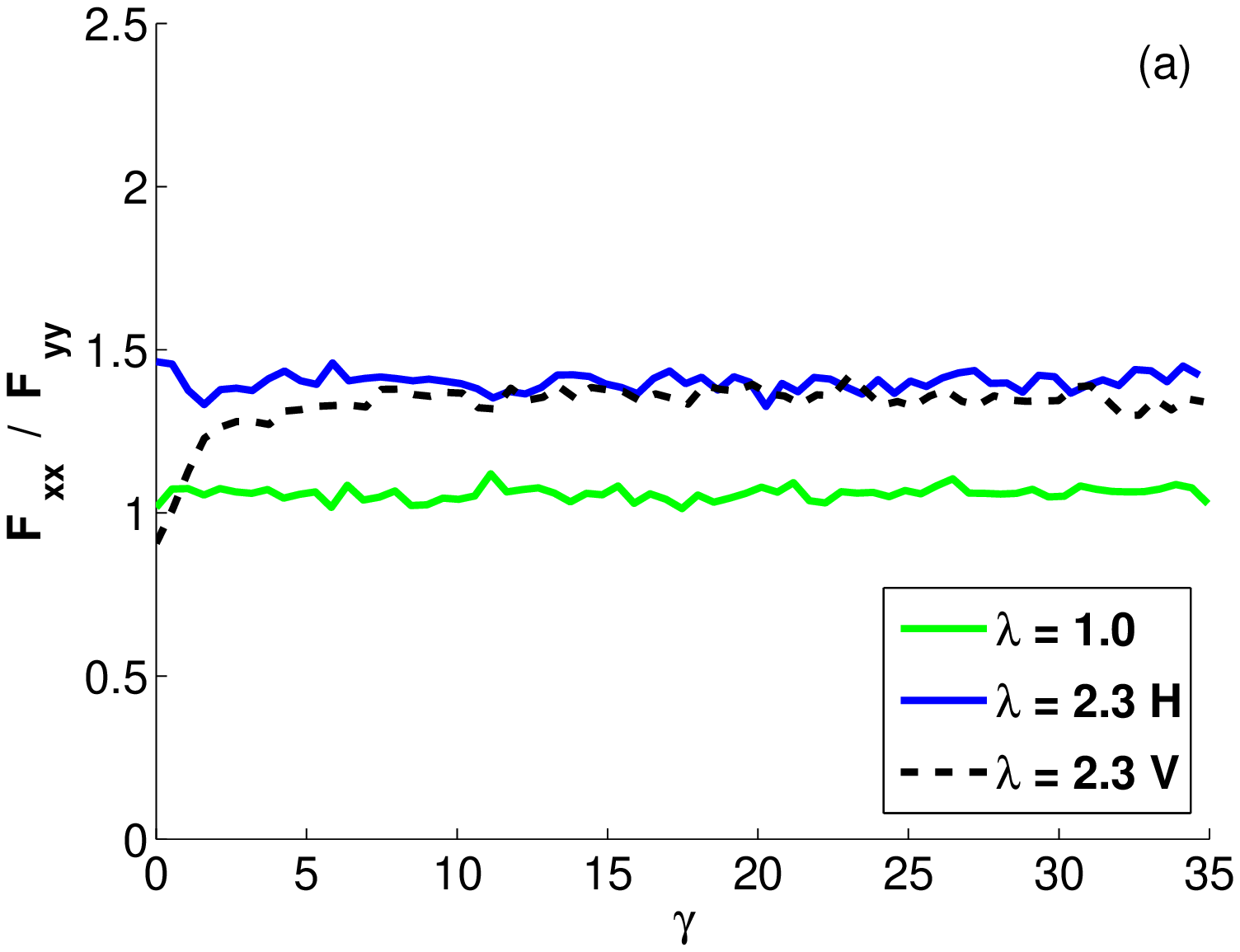,height=5.3cm,angle=0} \\
    \psfig{file=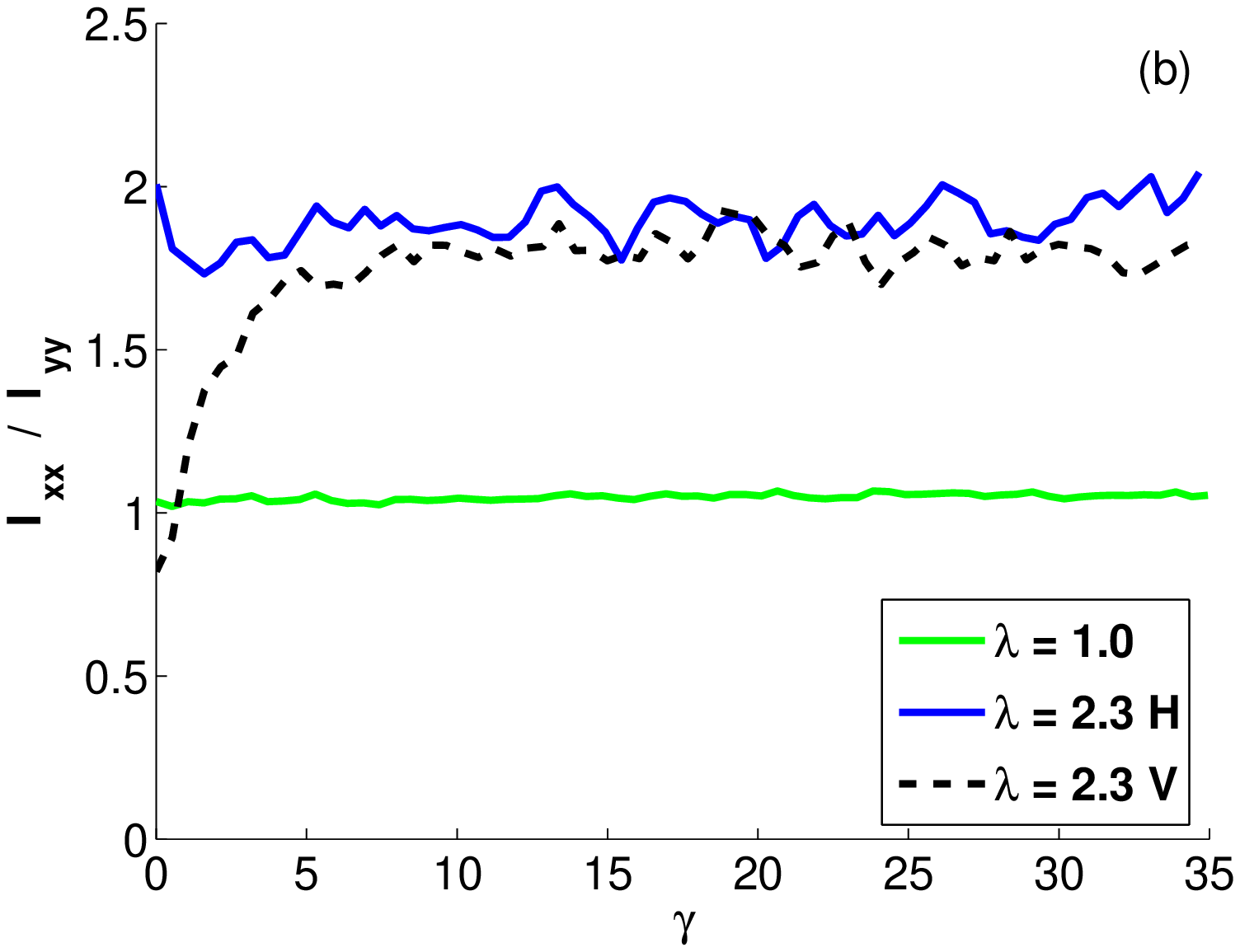,height=5.3cm,angle=0}
    \caption{Evolution of the quotient of the principal components of the fabric (a) and the
    inertia (b) tensors for isotropic particles  $\lambda$ = 1.0, and elongated particles ($\lambda$ =
    2.3)
    initially oriented in horizontal direction (H) and in vertical direction (V). } \label{Fig:quotT}
\end{figure}

The anisotropic distribution of
the contact orientations is more pronounced in case (b) than in
case (c); this is most probably due to the shape of the shear cell
(which indeed is wider than higher) and the compression process
using rigid walls. The orientational distribution of the contacts in the steady state
is depicted in Figure $\ref{Fig:Init-Steady-fabric}$(d-f). We notice that
the distribution of contact orientations for elongated particles (Fig. $\ref{Fig:Init-Steady-fabric}$(e-f))
is very similar independent of their initial orientation, while
for isotropic particles (Fig. $\ref{Fig:Init-Steady-fabric}$(d)) it is clearly different. The major
principal direction of the fabric tensor follows this same trend.

In Figure $\ref{Fig:ParticleOri}$, the polar distribution of
$\theta^p$ for elongated particles, and the principal directions
of the mean inertia tensor ($I_M$  and $I_m$) in the beginning and
in the stationary state are presented. We observe that, similar to
the case of contact orientations, they evolve towards the same
global orientation independently of the initial particle
directions.

In order to study the evolution of the fabric and inertia tensors
we monitor their deviatoric component $F_{yy} - F_{xx}$, and the
quotient $F_{xx}$ / $F_{yy}$ during the simulation. One can
observe in Figure $\ref{Fig:DevCompon}$, where the evolution of
the deviatoric component of $F_{ij}$ and $I_{ij}$ is shown, that
the deviatoric reaches a stationary value for both types of
particles, and that the induced anisotropy is much higher for
elongated particles than for isotropic ones. The same result is
observed for the quotient of the principal components of the
tensors (Figure $\ref{Fig:quotT}$). This stationary value of the
deviatoric component and the quotient is directly related to the
steady state at the macro-mechanical level, and seems to be a
micro-mechanical requirement for the global steady state. This
assumption is supported by simulations of biaxial tests reported
by Nouguier-Lehon et al. \cite{nouguier03}, where samples with
elongated particles do not reach neither a stationary value for
the components of the fabric and the orientation tensors nor the
so-called critical state at the macro-mechanical level, but
samples that reach the stationary state for the components of the
tensors do so at the global level.

\begin{figure}
\centering
    \psfig{file=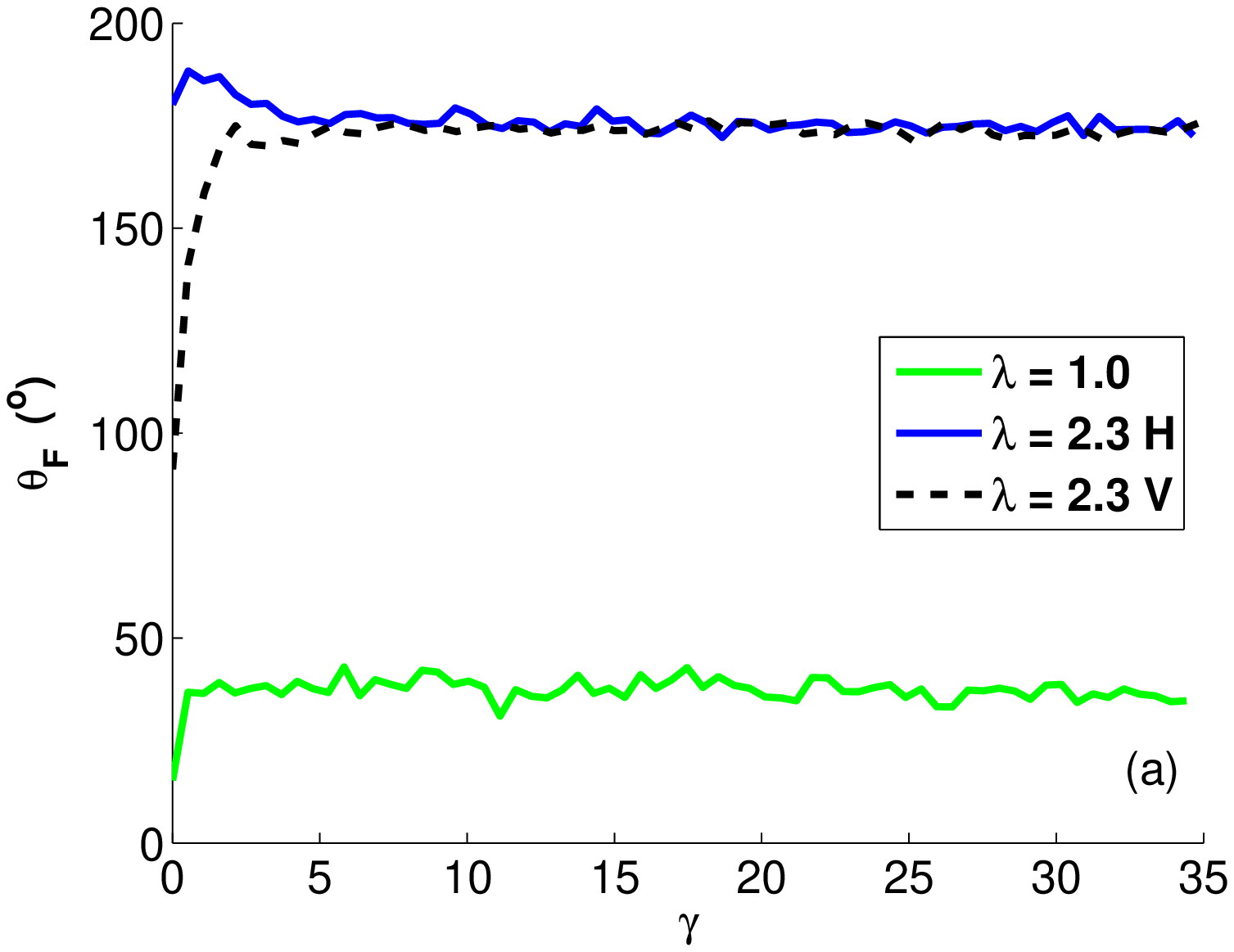,height=5.3cm,angle=0} \\
    \psfig{file=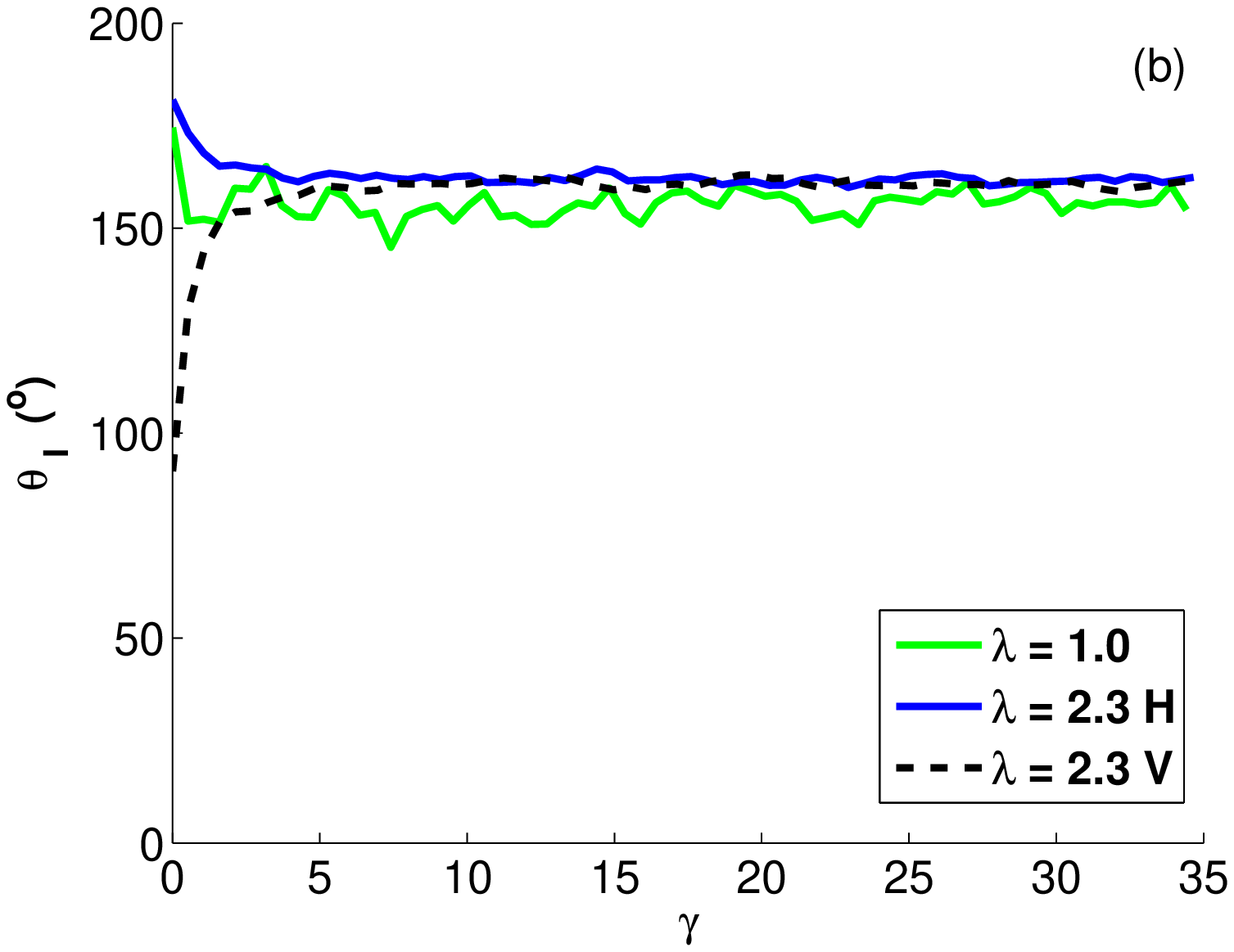,height=5.3cm,angle=0} \\
    \psfig{file=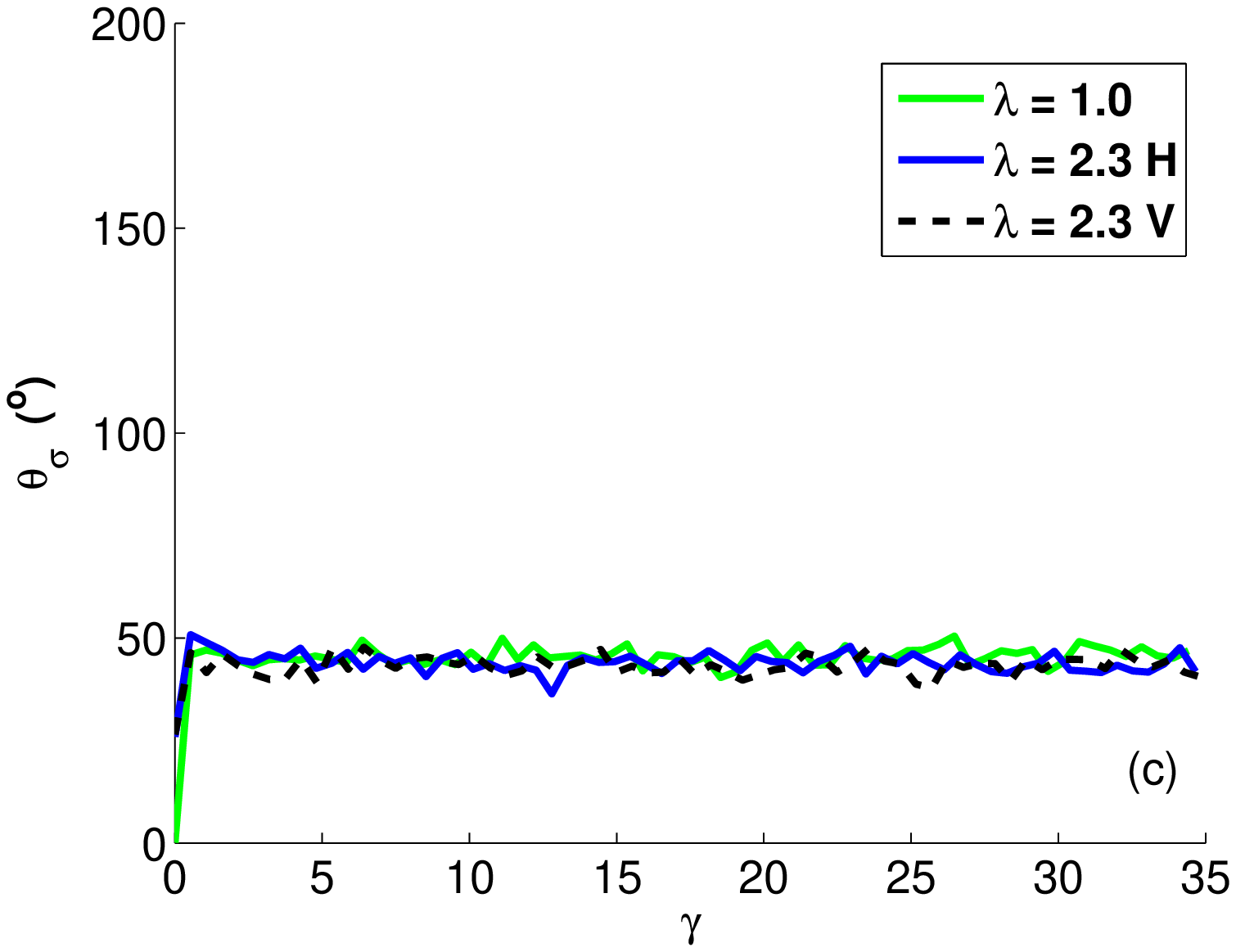,height=5.3cm,angle=0}
    \caption{Evolution of the major principal direction of the fabric (a), the inertia (b), and
    the stress (c) tensors  for isotropic particles  $\lambda$ = 1.0, and elongated particles ($\lambda$ =
    2.3)
    initially oriented in horizontal direction (H) and in vertical direction (V). } \label{Fig:PrincipalDirection}
\end{figure}

\begin{figure*}
\centering
    \psfig{file=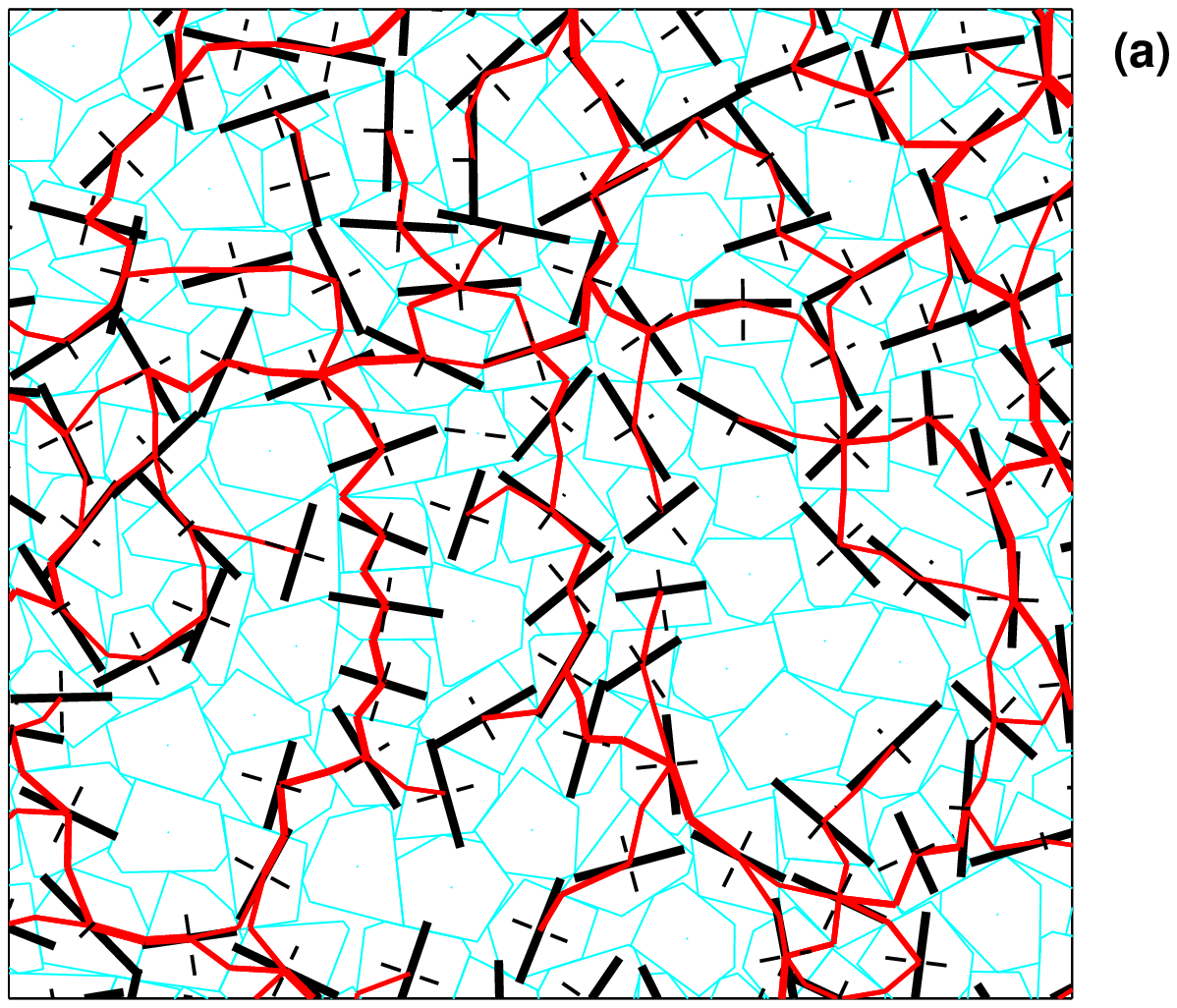,height=5.5cm,angle=0}
    \psfig{file=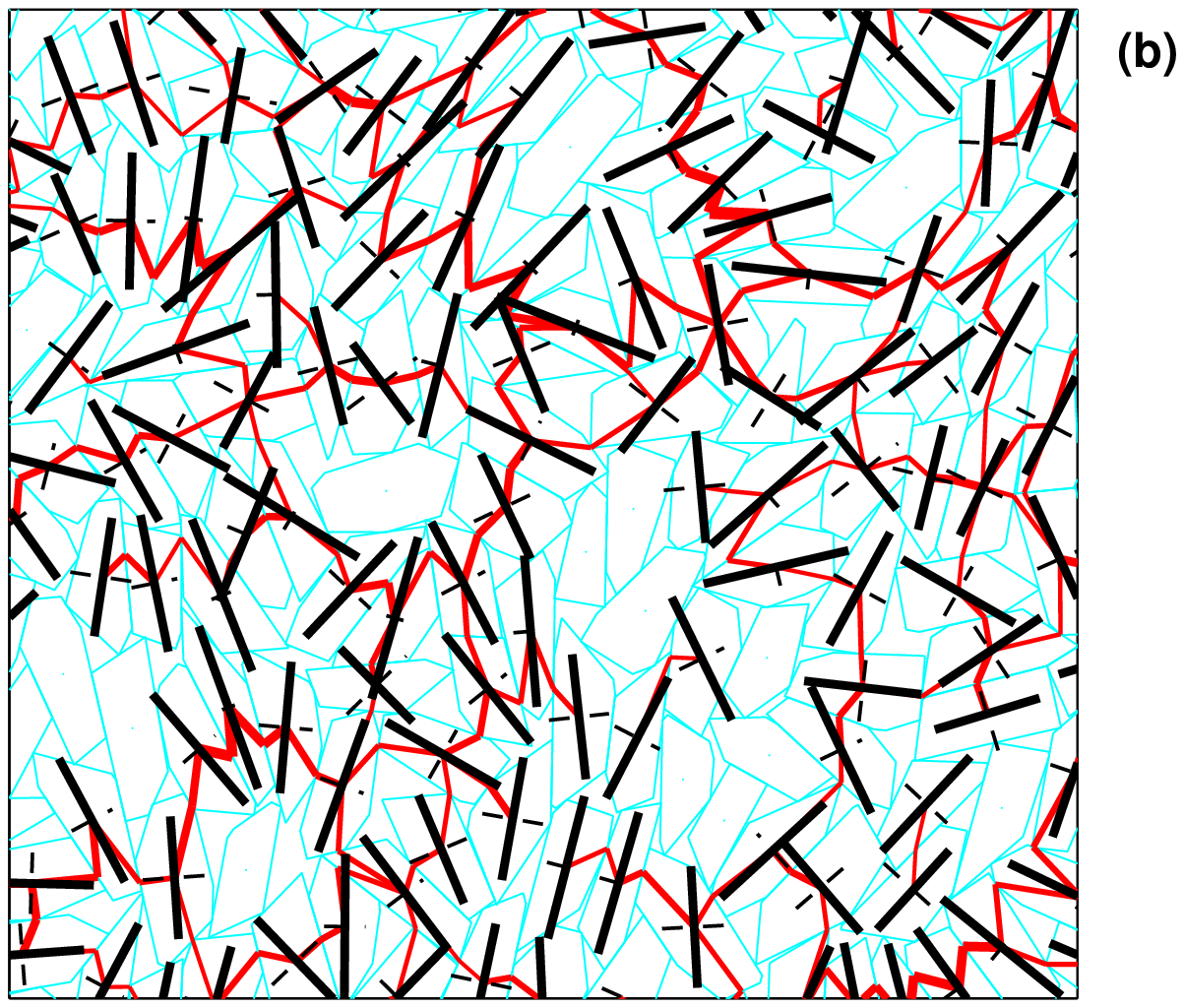,height=5.5cm,angle=0} \\
    \psfig{file=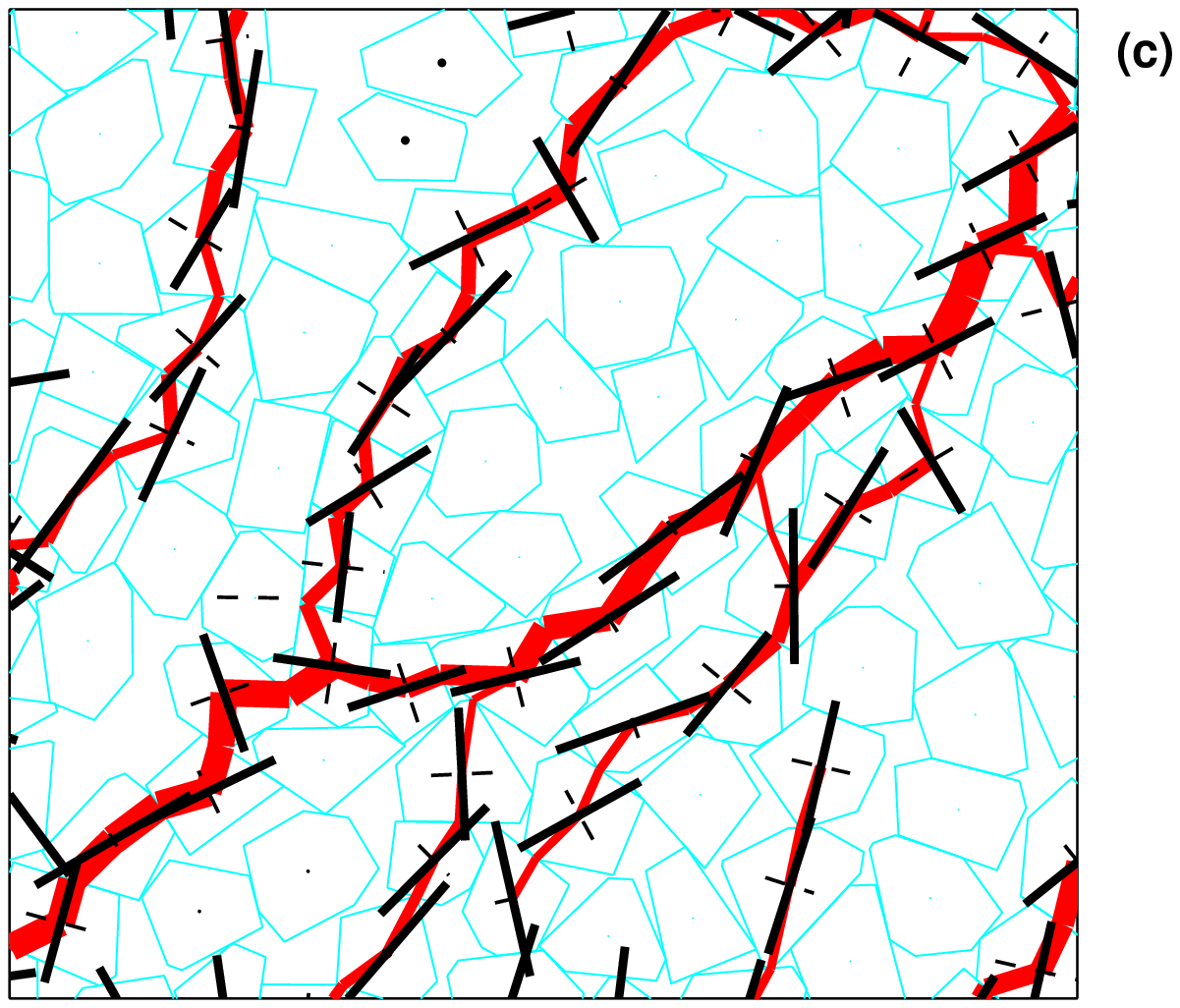,height=5.5cm,angle=0}
    \psfig{file=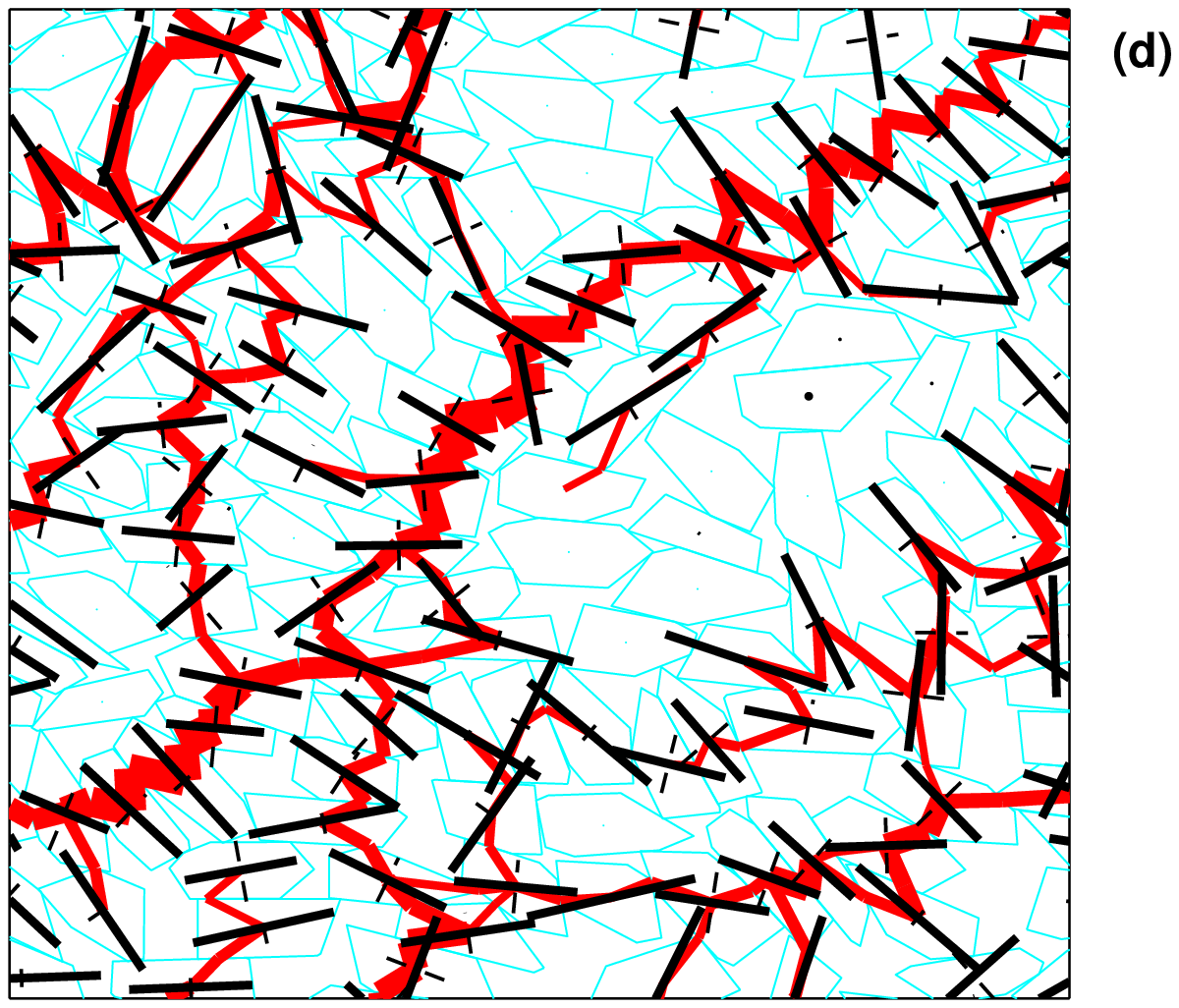,height=5.5cm,angle=0}
    \caption{Force chains (light lines, thickness proportional to magnitude) and principal axes of the fabric tensor (black lines),
     for initial configuration (a,b) and the steady state (c,d), for isotropic $\lambda$ = 1.0 (a,c)
     and elongated particles  $\lambda$ = 2.3 (b,d). } \label{Fig:DetailFIF}
\end{figure*}

Furthermore, for samples with elongated polygons the deviatoric
part of the $F_{ij}$ and $I_{ij}$ tensors, and the ratio of their
principal components reach approximately the same stationary value
independent of the initial particle orientations. This means that
the initial inherent anisotropy (fabric and particle orientation)
is completely erased and reoriented in direction of the induced
shear during the experiment. The evolution of the major principal
direction of the fabric $\theta_F$, the inertia $\theta_I$ and the
stress tensors $\theta_\sigma$ are shown in Figure
$\ref{Fig:PrincipalDirection}$. In the case of the inertia tensor,
the major principal direction $\theta_I$ is reoriented for all
samples towards an angle close to $160^\circ$. For the stress
tensor, $\theta_\sigma$ ($\approx$ $45^\circ$) is the same for
both particles independent of the particle shape. This orientation
of the stress comes from the direction of the force chains
carrying the largest stresses (Fig. $\ref{Fig:DetailFIF}$(c-d)).
On the other hand, the major principal direction of the
fabric tensor $\theta_F$ is completely different for isotropic and
elongated polygons. For isotropic particles it is similar to the
major principal direction of stress $\theta_\sigma$, but for
elongated polygons the fabric orientation is close to the one of
the inertia tensor.

To clarify this result further, the contact forces larger than the
mean value and the principal axes of the fabric tensor of the
corresponding particles are plotted
in Figure $\ref{Fig:DetailFIF}$. This is done for both types of polygons, and for the
initial configuration and a snapshot in the steady state. In Figure
$\ref{Fig:DetailFIF}$(a), where the initial configuration of
isotropic particles is shown, one can observe that the major
principal axis of the fabric tensor of each particle $F_{M}^p$ is
oriented independently of the orientation of the force chains. In
the system with elongated particles, however, $F_{M}^p$ is
slightly oriented in the largest dimension of the particles (major
principal axis of the tensor of inertia of each particle
$i_{M}^p$).

In the stationary state, we notice that in the system with
isotropic particles $F_{M}^p$ approximately follows the direction
of the force chains that carry the larger forces. In the one with
elongated particles, on the contrary, $F_{M}^p$ is oriented in
direction of the largest dimension of the particles $i_{M}^p$. The orientation of the elongated
particles within the main force chains is associated to the
stability of the packing. That is to say; forces are transmitted
through contacts closely parallel to the minor principal axis of
the inertia tensor of each particle $i_{m}^p$, in which flat
contact surfaces give a more stable configuration to the system.
We conclude then that the orientation of the contacts in the steady
state, in the case of non-spherical particles is governed by the
particle orientation, and for isotropic particles by the direction
of the major principal stress. This is also observed in Figure
$\ref{Fig:PrincipalDirection}$, where in the steady state for
isotropic particles $\theta_F$ is almost the same as
$\theta_\sigma$, and for elongated particles $\theta_F$ is nearly
$\theta_I$.

\subsection{Shear localization and particle rotation}

\begin{figure}
\centering
    \psfig{file=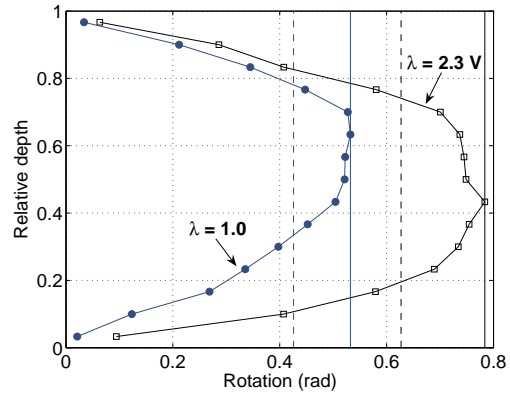,height=5.5cm,angle=0}
    \caption{Average accumulated rotation of the particles during the steady state,
    isotropic $\lambda$ = 1.0 (full dots) and elongated
    particles $\lambda$ = 2.3 V (open squares) within horizontal layers as a function of relative depth.
    The maximum and 80 $\%$ of the
    maximum value of rotation are plotted as solid and dashed lines, respectively.}
    \label{Fig:Localization}
\end{figure}

In order to study strain localization and particle rotation, the
shear cell is divided into horizontal layers, i.e parallel to
shear direction. For a clearer presentation of the results, we
normalize the vertical dimensions with the height of the system
$h$. The origin corresponds to the bottom and 1 to the top of the
sample. We use in our analysis the rotation that particles
accumulate during every unit increment ($\Delta \gamma_{unit}$) of
the strain variable $\gamma$ in the steady state (in our
experiment we take $\gamma_{initial}$ = 10 and $\gamma_{final}$ =
35, i.e. in total 25 $\Delta \gamma_{unit}$). Then, we average
this accumulated particle rotation for each layer of the system,
and for all the considered strain increments. In Figure
$\ref{Fig:Localization}$, the average accumulated rotation in the
steady state for each layer and a shear velocity of 40 cm/s for
isotropic and elongated particles initially oriented in vertical
direction is shown. We observe a clear localization of rotations,
having a peak close to the center and decreasing toward the
boundaries. This distribution resembles the movement
of two rigid bodies against each other on a shear band.

We are also interested in the width of the shear zone, which we
define here as the width of the zone with particle rotation larger
than 80 $\%$ of the maximum rotation. One can notice that there
are two important differences between the two types of particles
considered:

\begin{itemize}
    \item The average accumulated rotation is lower for elongated particles, in this
    particular case it is 65 $\%$ of the isotropic particles rotation.
    \item The width of the localization zone is smaller for
    elongated particles (0.45 times the system height $h$ for elongated and 0.55 times $h$ for
    isotropic particles).
\end{itemize}

\begin{figure}
\centering
    \psfig{file=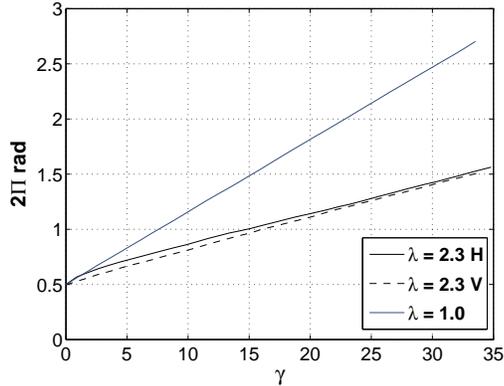,height=5.5cm,angle=0}
    \caption{Mean accumulated rotation
     since the beginning of the simulation for isotropic particles $\lambda$ = 1.0 (light line),
     and elongated particles ($\lambda$ = 2.3) initially oriented in horizontal direction (H, black line), and
     in vertical direction (V, black-dashed line).  }
    \label{Fig:Rotat}
\end{figure}

\begin{figure}
\centering
    \psfig{file=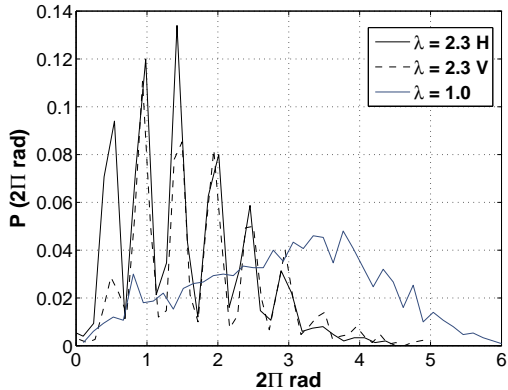,height=5.5cm,angle=0}
    \caption{Probability distribution function of accumulated rotation
     since the beginning of the simulation until $\gamma$ = 35, for isotropic particles ($\lambda$ = 1.0, light line),
     and elongated particles ($\lambda$ = 2.3) initially oriented in horizontal direction (black line), and
     in vertical direction (black-dashed line). }
    \label{Fig:PDF}
\end{figure}

These differences in accumulated rotation and relative width
between isotropic and elongated particles can be explained by the
frustration of movement and rotation that elongated particles
experience due to the stronger interlocking among them. In this
way, the localization zone (rotation zone) for elongated polygons
becomes thinner than for isotropic ones. In Figure
$\ref{Fig:Rotat}$ the mean accumulated rotation since the
beginning of the simulation for isotropic and elongated particles
is depicted. Notice that the mean rotation is almost twice for
isotropic than for elongated particles at the end of the
simulation. We also calculate the probability distribution
function of particle rotation, which is shown in Figure
$\ref{Fig:PDF}$ for both isotropic and elongated particles and
$\gamma$ = 35. For isotropic particles a more uniform distribution
is observed, and the maximum value is close to four complete
rotations (a complete rotation $2\pi \;rad$).  For elongated particles
the probability
distribution function presents several peaks every half
of
rotation ($\pi \;rad$). This fact indicates the strong frustration of rotation that such particles undergo during shearing, and that
the
typical mode of accumulating rotation is then every half complete rotation.

\section{Discussion and final remarks}
\label{discussion}

In this paper, the influence of particle shape on the mechanical
behavior of sheared dense granular media has been investigated
using two-dimensional discrete numerical simulations. Our results
show an important influence of particle anisotropy on both the
macro and micro-mechanical behavior of the granular media. It is
found that for samples with isotropic and elongated particles the
shear force and volumetric strain saturate at constant values
reaching a steady state. These values in the case of elongated
particles are higher than for isotropic particles due to the
stronger interlocking between anisotropic particles. This is
consequence of the geometrical effect of particle shape.
Furthermore, samples with elongated particles reach the same
saturation value in the steady state independently of the initial
orientation of the particles. This is related to the removal and
reorientation of the initial inherent anisotropy (fabric and
particle orientations) in direction of the induced shear. This has
been confirmed by studying the evolution of the fabric and the
inertia tensors.

The deviatoric part and the quotient of the principal components
of the fabric $\mathbf{F}$ and inertia $\mathbf{I}$ tensors, for
both types of particles, reach a stationary value independent of
their initial one. This is directly related to the steady state at
the macro-mechanical level. The principal directions of
$\mathbf{F}$ and $\mathbf{I}$ present the following behavior: in
the initial state of the samples, isotropic polygons present no
preferred direction of contacts, however, in the case of elongated
polygons the major principal component of $\mathbf{F}$ is oriented
along the direction of the major principal component of
$\mathbf{I}$. In the steady state, ${F_M}$ in the case of
isotropic particles is reoriented in the direction of the major
component of the stress tensor $\mathbf{\sigma}$, but for
elongated particles ${F_M}$ evolves following the induced
orientation that particles undergo during shearing. The direction
of the major component of $\mathbf{\sigma}$ is the same for both
particle shapes. Independently of the initial orientation, samples
with elongated particles reach the same contact $\theta_F$ and
particle $\theta_I$ global orientation in the steady state. One
can then conclude that a stationary value of the principal
components and principal directions of the fabric and inertia
tensors is a micro-mechanical requirement for the existence of the
global steady state of the medium. We also conclude that for the
sheared dense granular materials here presented, the contact
orientation in the global stationary state is governed in the case
of isotropic particles by the direction of the major principal
component of the stress tensor, and for elongated particles mainly
by the major principal component of the inertia tensor (particle
orientation).

At the particle level, these results are clearly understood when
the inertia and fabric tensor of the particles within the force
chains carrying the larger forces are studied. We found that the
orientation of elongated particles is associated to the stability
of the packing, i.e. forces are transmitted through contacts
parallel to the shortest dimension of the particles $i_{m}^p$.

Concerning strain localization and particle rotation the following
is observed: the width of the shear zone and the accumulated
rotation is larger for isotropic particles than for elongated
particles. This results is explained by the frustration of
rotation that elongated particles experience due to the stronger
interlocking among them. This is clearly observed in the
probability distribution function of particle rotation, where the
typical mode of accumulating rotation for elongated particles is
every $\pi \;rad$.

Results presented in this paper lead to a better comprehension of
the role of particle shape on the macro and micro-mechanical
response of granular materials. It is, however, necessary to say
that due to the nature of our two-dimensional analysis, a
validation with a three-dimensional model would be very helpful.
Further work will focus on a systematic study of the influence of
different aspect ratios of the particles, beyond the two limit
cases here studied, on the observed phenomena. Additionally, the
influence of two types of particles (isotropic and anisotropic) as
constituents of the sample on the mechanical behavior of the
granular media will be investigated.

\begin{acknowledgements}
The authors would like to thank F. Alonso-Marroqu\'{\i}n for
helpful discussions.  They also want to acknowledge the support of
the \textit{Deutsche Forschungsgemeinschaft} Project HE 2732781
\textit{Micromechanische Untersuchung des granulares Ratchetings}
and the EU project Degradation and Instabilities in Geomaterials
with Application to Hazard Mitigation (DIGA) in the framework of
the Human Potential Program, Research Training Networks
(HPRN-CT-2002-00220).
\end{acknowledgements}


\begin{thebibliography}{3}
\providecommand{\natexlab}[1]{#1}
\providecommand{\url}[1]{\texttt{#1}}
\expandafter\ifx\csname urlstyle\endcsname\relax
  \providecommand{\doi}[1]{doi: #1}\else
  \providecommand{\doi}{doi: \begingroup \urlstyle{rm}\Url}\fi

\bibitem[Casagrande and Carillo(1944)]{Casag44}
A.~Casagrande and N.~Carillo.
\newblock Shear failure of anisotropic materials.
\newblock In \emph{Proc. Boston Society of Civil Engrs 31.}, pages 74--78,
  1944.

\bibitem[Oda(1972)]{Oda72}
M.~Oda.
\newblock Initial fabrics and their relations to mechanical properties of
  granular materials.
\newblock \emph{Soils Found.}, 12\penalty0 (1):\penalty0 17--36, 1972.

\bibitem[Oda et~al.(1985)Oda, Nemat-Nasser, and Konishi]{Oda85}
M.~Oda, S.~Nemat-Nasser, and J.~Konishi.
\newblock Stress-induced anisotropy in granular masses.
\newblock \emph{Soils Found.}, 25\penalty0 (3):\penalty0 85--97, 1985.

\bibitem[Oda and Nakayama(1988)]{Oda88}
M.~Oda and H.~Nakayama.
\newblock Introduction of inherent anisotropy of soils in the yield function.
\newblock In M.~Satake and J.~T. Jenkins, editors, \emph{Micromechanics of
  granular materials}, pages 81--90. Elsevier, 1988.

\bibitem[Li and Dafalias(2002)]{Li02}
X.~S. Li and Y.~F. Dafalias.
\newblock Constitutive modeling of inherently anisotropic sand behavior.
\newblock \emph{J. Geotech. Geoenviron. Engng.}, 128\penalty0 (10):\penalty0
  868--880, 2002.

\bibitem[Bowman et~al.(2001)Bowman, Soga, and Drummond]{bowman01}
E.~T. Bowman, K.~Soga, and W.~Drummond.
\newblock Particle shape characterization using fourier descriptor analysis.
\newblock \emph{G\'eotechnique}, 51\penalty0 (6):\penalty0 545--554, 2001.

\bibitem[Matsushima and Saomoto(2002)]{matsuhima02}
T.~Matsushima and H.~Saomoto.
\newblock Discrete element modeling for irregularly-shaped sand grains.
\newblock In Mestat, editor, \emph{Proc. NUMGE2002: Numerical Methods in
  Geotechnical Engineering}, pages 239--246, 2002.

\bibitem[Bowman and Soga(2005)]{bowman05}
E.T Bowman and K.~Soga.
\newblock The influence of particle shape on the stress-strain and creep
  response of fine silica sand.
\newblock In R.~Garc\'{\i}a-Rojo, H.J. Herrmann, and S.~McNamara, editors,
  \emph{Powders and Grains 2005}, pages 1325--1328. Balkema, 2005.

\bibitem[Shodja and Nezami(2003)]{shodja03}
H.~M. Shodja and E.~G. Nezami.
\newblock A micromechanical study of rolling and sliding contacts in assemblies
  of oval granules.
\newblock \emph{Int. J. Numer. Anal. Meth. Geomech.}, 27:\penalty0 403--424,
  2003.

\bibitem[Nouguier-Lehon and Frossard(2005)]{nouguier05}
C.~Nouguier-Lehon and E.~Frossard.
\newblock Influence of particle shape on rotations and rolling movements in
  granular media.
\newblock In R.~Garc\'{\i}a-Rojo, H.J. Herrmann, and S.~McNamara, editors,
  \emph{Powders and Grains 2005}, pages 1339--1343. Balkema, 2005.

\bibitem[Nouguier-Lehon et~al.(2003)Nouguier-Lehon, Cambou, and
  Vincens]{nouguier03}
C.~Nouguier-Lehon, B.~Cambou, and E.~Vincens.
\newblock Influence of particle shape and angularity on the behavior of
  granular materials: a numerical analysis.
\newblock \emph{Int. J. Numer. Anal. Meth. Geomech.}, 27:\penalty0 1207--1226,
  2003.

\bibitem[Pe{\~n}a et~al.(2005)Pe{\~n}a, Lizcano, Alonso-Marroqu\'{\i}n, and
  Herrmann]{pena05}
A.~A. Pe{\~n}a, A.~Lizcano, F.~Alonso-Marroqu\'{\i}n, and H.~J. Herrmann.
\newblock Investigation of the asymptotic states of granular materials using a
  discrete model of anisotropic particles.
\newblock In R.~Garc\'{\i}a-Rojo, H.J. Herrmann, and S.~McNamara, editors,
  \emph{Powders and Grains 2005}, pages 697--700. Balkema, 2005.

\bibitem[Ng(2001)]{ng01}
T.~T. Ng.
\newblock Fabric evolution of ellipsoidal arrays with different particle
  shapes.
\newblock \emph{ASCE J. Engrg. Mech.}, 127:\penalty0 994--99, 2001.

\bibitem[Ng(2004)]{ng04}
T.~T. Ng.
\newblock Behavior of ellipsoids of two sizes.
\newblock \emph{J. Geotech. Geoenviron. Engng, ASCE}, 130(10):\penalty0
  1077--1083, 2004.

\bibitem[Villarruel et~al.(2000)Villarruel, Lauderdale, Mueth, and
  Jaeger]{villarruel00}
Fernando~X. Villarruel, Benjamin~E. Lauderdale, Daniel~M. Mueth, and
  Heinrich~M. Jaeger.
\newblock Compaction of rods: relaxation and ordering in vibrated, anistopic
  granular material.
\newblock \emph{Phys. Rev. E}, 61:\penalty0 6914, 2000.

\bibitem[Lumay and Vandewalle(2004)]{lumay04}
G.~Lumay and N.~Vandewalle.
\newblock Compaction of anisotropic granular materials.
\newblock \emph{Phys. Rev. E}, 70:\penalty0 051314, 2004.

\bibitem[Ribi\`ere et~al.(2005)Ribi\`ere, Richard, Bideau, and
  Delannay]{ribiere05}
P.~Ribi\`ere, P.~Richard, D.~Bideau, and R.~Delannay.
\newblock Experimental compaction of anisotropic granular media.
\newblock \emph{The European Physical Journal E}, 16:\penalty0 415--420, 2005.

\bibitem[Majmudar and Behringer(2005)]{majmudar05}
T.~S. Majmudar and R.~P. Behringer.
\newblock Contact force measurements and stress-induced anisotropy in granular
  materials.
\newblock \emph{Nature}, 435\penalty0 (23):\penalty0 1079--1082, 2005.

\bibitem[L\"atzel et~al.(2000)L\"atzel, Luding, and Herrmann]{latzel00}
M.~L\"atzel, S.~Luding, and H.~J. Herrmann.
\newblock Macroscopic material properties from quasi-static, microscopic
  simulations of a two-dimensional shear-cell.
\newblock \emph{Granular Matter}, 2\penalty0 (3):\penalty0 123--135, 2000.

\bibitem[Tillemans and Herrmann(1995)]{tillemans95}
H.-J. Tillemans and H.~J. Herrmann.
\newblock Simulating deformations of granular solids under shear.
\newblock \emph{Physica A}, 217:\penalty0 261--288, 1995.

\bibitem[Kun and Herrmann(1996)]{kun96b}
F.~Kun and H.~J. Herrmann.
\newblock A study of fragmentation processes using a discrete element method.
\newblock \emph{Comput. Methods Appl. Mech. Eng.}, 138:\penalty0 3--18, 1996.

\bibitem[Kun and Herrmann(1999)]{kun99}
Ferenc Kun and Hans~J. Herrmann.
\newblock Transition from damage to fragmentation in collision of solids.
\newblock \emph{Phys. Rev. E}, 59\penalty0 (3):\penalty0 2623--2632, 1999.

\bibitem[Alonso-Marroquin and Herrmann(2002)]{alonso02}
F.~Alonso-Marroquin and H.J. Herrmann.
\newblock Calculation of the incremental stress-strain relation of a polygonal
  packing.
\newblock \emph{Phys. Rev. E}, 66:\penalty0 021301, 2002.

\bibitem[Veje et~al.(1999)Veje, Howell, and Behringer]{veje99}
C.~T. Veje, D.~W. Howell, and R.~P. Behringer.
\newblock Kinematics of a {2D} granular {C}ouette experiment at the transition
  to shearing.
\newblock \emph{Phys. Rev. E}, 59:\penalty0 739, 1999.

\bibitem[Howell et~al.(1999)Howell, Behringer, and Veje]{howell99}
D.~Howell, R.~P. Behringer, and C.~Veje.
\newblock Stress fluctuations in a 2d granular {C}ouette experiment: {A}
  continuous transition.
\newblock \emph{Phys. Rev. Lett.}, 82:\penalty0 5241, 1999.

\bibitem[Moukarzel and Herrmann(1992)]{moukarzel92}
C.~Moukarzel and H.~J. Herrmann.
\newblock A vectorizable random lattice.
\newblock \emph{Journal of Statistical Physics}, 68:\penalty0 911--923, 1992.

\bibitem[Rothenburg and Selvadurai(1981)]{rothenburg81}
L.~Rothenburg and A.~P.~S. Selvadurai.
\newblock A micromechanical definition of the cauchy stress tensor for
  particulate media.
\newblock In A.~P.~S. Selvadurai, editor, \emph{Mechanics of Structured Media},
  pages 469--486. Elsevier, 1981.

\end{thebibliography}
\end{document}